\documentclass[ba]{imstartnew}
\pubyear{2024+}
\volume{TBA}
\issue{TBA}
\firstpage{1}
\lastpage{1}
\usepackage{rotating}
\usepackage{amsthm}
\usepackage{amsmath}
\usepackage{natbib}
\usepackage[colorlinks,citecolor=black,urlcolor=black,filecolor=black,backref=page]{hyperref}
\usepackage{graphicx}

\usepackage{amsmath}
\usepackage{graphicx}
\usepackage{enumerate}
\usepackage{pdflscape}

\usepackage{url} 
\usepackage[utf8]{inputenc} 
\usepackage{graphicx}
\usepackage{xcolor,yfonts,eufrak}
\usepackage{blkarray}
\usepackage{amsmath,amssymb,amsthm,amsfonts,natbib,bbm,bm,color, graphics,graphicx,indentfirst,url,multicol,multirow,longtable,tabu,verbatim,enumerate,footnote,multirow}
\usepackage{tikz,tikzscale}
\usepackage{pifont}
\usetikzlibrary{fit,positioning}
 \usetikzlibrary{bayesnet, arrows}
\usepackage{graphicx}
 \usepackage{multirow}
 \usepackage{lscape}
\usepackage{tablefootnote}
\usepackage{ulem}
\usepackage{amssymb}
\usepackage{algorithm}
\usepackage{algpseudocode}
\usepackage{xcolor}
\usepackage{soul}
\usepackage{amsmath}
\usepackage{enumerate}
\usepackage{natbib} 

\usepackage{tikz,tikzscale}
\usepackage{pifont}
\usetikzlibrary{fit,positioning}
 \usetikzlibrary{bayesnet}

\startlocaldefs
\endlocaldefs

\begin{document}


\begin{frontmatter}
\title{A Unified Bayesian Framework for Modeling Measurement Error in Multinomial Data}
\runtitle{Measurement Error in Multinomial Data}

\begin{aug}
\author{ \snm{Matthew D. }\fnms{Koslovsky}\thanksref{addr1}\ead[label=e1]{\\matt.koslovsky@colostate.edu}},
\author{\fnms{Andee} \snm{Kaplan}\thanksref{addr3}\ead[label=e3]{ \\andee.kaplan@colostate.edu}},
\author{\fnms{Victoria A.} \snm{Terranova}\thanksref{addr4}\ead[label=e4]{victoria.terranova@unco.edu}}, \\
\and
\author{\snm{Mevin B.} \fnms{Hooten}\thanksref{addr2,t1,m2}\ead[label=e2]{mevin.hooten@austin.utexas.edu}}

\runauthor{Koslovsky, M.D. et al.}

\address[addr1]{Department of Statistics, Colorado State University, Fort Collins, CO, USA.
    \printead{e1} 
}  

\address[addr3]{Department of Statistics, Colorado State University, Fort Collins, CO, USA.
    \printead{e3} 
}  

\address[addr4]{Department of Criminology \& Criminal Justice, University of Northern Colorado, Greeley, CO, USA.
    \printead{e4} 
}  

\address[addr2]{Department of Statistics and Data Sciences, The University of Texas at Austin, Austin, TX, USA. \printead{e2} 
} 

\end{aug} 

\begin{abstract}
Measurement error in multinomial data is a well-known and well-studied inferential problem that is encountered in many fields, including engineering, biomedical and omics research, ecology, finance, \textcolor{black}{official statistics}, and social sciences.  Methods developed to accommodate measurement error in multinomial data are typically equipped to handle false negatives or false positives, but not both. We provide a unified framework for accommodating both forms of measurement error using a Bayesian hierarchical approach. We demonstrate the proposed method's performance on simulated data and apply it to acoustic bat monitoring \textcolor{black}{and official crime} data.
\end{abstract}


\end{frontmatter}

\section{Introduction}

Measurement error in multinomial data 
is a well-known and well-studied inferential problem that is encountered in many fields, including engineering, biomedical and omics research, ecology, finance, \textcolor{black}{official statistics}, and social sciences \citep{swartz2004bayesian, perez2007misclassified, molinari2008partial, datta2021regularized, mulick2022bayesian}.  In this work, we define measurement error as the discrepancy between an observed or measured variable and its true value and consider two types of measurement error defined as (1) false negatives that occur when a particular category or class is present in the population but it is not observed in the sample and (2) false positives that occur when a sampled observation is misclassified into the wrong category. While both types of measurement error may be present in the data, existing methods are not designed to accommodate them simultaneously which may bias inference. \textcolor{black}{To fill this gap, we propose a unified framework for accommodating both forms of measurement error when modeling multinomial data. } \textcolor{black}{Our approach differs from existing methods in that it  models the probability of misclassification for each individual observation explicitly; is scalable to high-dimensional classification problems; and accommodates individual-level covariates associated with the probability of being a true/false negative, the true classification probabilities, and the probability of misclassification. }

 Modeling false negatives in multinomial data is closely related to the notion of handling zero-inflation in univariate and multivariate count data. Count data are considered zero-inflated when the number of zeroes observed in the data set is larger than expected under the assumptions of the sampling distribution. Zero-inflated count models are typically constructed as a two-component mixture of a point mass at zero and a sampling distribution for the count data (e.g., Poisson or negative binomial distributions in the univariate setting; \cite{xu2015assessment, zhang2020nbzimm,jiang2021bayesian,shuler2021bayesian}). To achieve this, an at-risk indicator is introduced into the model to differentiate between at-risk zeros (i.e., a zero count is observed, but there is a positive probability of occurrence) and structural zeros (i.e., a zero count is observed because there is zero probability of occurrence; \cite{neelon2019bayesian}), or equivalently between false negatives and true negatives, respectively. In multivariate settings, researchers link zero-inflated univariate count models via latent parameters that control the dependence structure between counts \citep{aitchison1989multivariate, chiquet2021poisson}. This approach models the multivariate counts unconditionally on the total count for the sample and is therefore not designed for multinomial classification problems or multivariate compositional count data where the total number of counts is fixed.  \cite{koslovsky2023bayesian} introduced a zero-inflated Dirichlet-multinomial (DM) distribution for handling excess zeros in multivariate compositional count data, which differs from traditional approaches for modeling zero-inflation in count data by assuming a mixture distribution on the count probabilities as opposed to the sampling distribution. Using a combination of data augmentation strategies, their approach is scalable to large compositional spaces, can accommodate covariates associated with zero-inflation and relative abundances, and has shown promising estimation performance in simulation. 

Existing methods designed to model false positives in multinomial data typically assume the observed classifications follow a multinomial distribution given the true (latent) classification \citep{swartz2004bayesian,wang2020inference,perez2007misclassified,frenay2013classification}. With this approach, the number of rows for the resulting matrix of classification probabilities equals the number of true categories, and the number of columns equals the number of observed categories with each row summing to one. The task of modeling false positives in multinomial data draws parallels to a popular approach for entity resolution. Entity resolution is the process of resolving duplicates in many overlapping data sets without the benefit of a unique identifying attribute. In the hit-miss approach to entity resolution, observed records are assumed to either represent the true records associated with an entity (hit) or a distorted version of this truth (miss) 
\citep{tancredi2011hierarchical,copas1990record}.  These potentially noisy records are then directly modeled using a mixture model in which two records that are associated with the same latent truth refer to the same entity and can be deduplicated \citep{steorts2016bayesian}. This direct approach of modeling measurement error in the likelihood is an analog to the misclassification problem we are interested in addressing in that the observed classifications can either be the true value (hit) or a distorted version of that truth (miss). Potential benefits of directly modeling the distortion process in a hit-miss framework include the ability to choose an appropriate distribution for the misclassification, incorporation of expert knowledge into the model via the priors on misclassification probabilities, and inference on the probability of misclassification after obtaining data.
 
 A fundamental issue shared by any model designed to accommodate misclassification is that the model is not identifiable without additional information about the true classification, as well as zero-inflation, process beyond the raw data \citep{swartz2004bayesian}. \textcolor{black}{Existing methods designed to accommodate false positives deal with identifiability using informative priors in Bayesian settings, auxiliary/calibration data to  estimate the matrix of classification probabilities (typically referred to as a confusion matrix) separately from the count model, or validating
the true classification of a subset of the data \citep{wright2020modelling, chambert2015modeling, guillera2017dealing}. \cite{stratton2022coupling} explore these strategies rigorously in simulation.} \cite{swartz2004bayesian} provide an extensive discussion of identifiability issues in multinomial classification models and propose using constraints to break the symmetry of the model, similar to what is done to accommodate label switching in Bayesian mixture models \citep{jasra2005markov}. 

In this work, we propose a novel method for simultaneously modeling false positives and false negatives in multinomial data.  Specifically, we assume a zero-inflated DM distribution to accommodate potential false negatives in the underlying true classifications as well as potential overdispersion. We then introduce a latent hit-miss indicator to model misclassification that allows our approach to differentiate between true detections and detection by chance.  The proposed model belongs to the class of semi-supervised learning methods because it can incorporate any amount of individual-level validation data, including no validation data in unsupervised settings.  We use a combination of data augmentation techniques to scale the model to high-dimensional settings found in practice. 
\textcolor{black}{ In a variety of simulation settings, we demonstrate the improved estimation performance of the proposed method compared to alternative approaches for handling false positives or false negatives in multivariate count data.  We then apply  the proposed method to two data sets collected in ecological and criminal justice research. Applied to bat monitoring data, we show how the proposed method can serve as an alternative approach for accommodating imperfect detection in multispecies occupancy-detection modeling. Our approach differs from existing multispecies occupancy-detection models because it does not require specification of the ecological process and instead models the true latent classifications explicitly. In a second application study, we show how the proposed method can accommodate potential biases attributed to zero-inflation and misclassification in official crime data.  By taking a fully-Bayesian approach, our method propagates the uncertainty of potential false positives and false negatives in the estimation of parameters of interest.  }    

\section{Methods}

In this section, we present a general formulation for modeling measurement error in multinomial data, making connections to relevant occupancy-detection \textcolor{black}{and official crime data} modeling aspects as necessary. 
Let the $C$-dimensional vector $\boldsymbol{y}_{ijl}$ represent the observed classification for the $i$th  $(i= 1, \dots, N$) observation (or site/location\textcolor{black}{/jurisdiction}) at the $j$th $(j= 1,\dots, n_i)$ measurement (or visit\textcolor{black}{/survey}) for the $l$th $(l = 1, \dots, L_{ij})$ individual (or organism\textcolor{black}{/incident}), where $y_{ijlc} = 1$ indicates the observed individual was classified into the $c$th category (or species\textcolor{black}{/crime}) and 0 otherwise.  In general, the model does not require more than one measurement of each site (i.e., $n_i > 1$). However in various fields, including ecological monitoring, each site is typically visited multiple times to improve inference \citep{mackenzie2002estimating, lele2012dealing}. We let the $T$-dimensional vector $\boldsymbol{z}_{ijl}$ represent the individual's true classification (or species\textcolor{black}{/crime}), where $z_{ijlt} = 1$ indicates the individual truly belongs to the $t$th category and 0 otherwise.  For ease of presentation, we assume $T=C$ and that the ordering of the elements is the same in $\boldsymbol{y}_{ijl}$ and $ \boldsymbol{z}_{ijl}$ (i.e., $y_{ijlt} = z_{ijlt} = 1 $ indicates that the latent and observed categories are the same).

For each individual, we introduce a latent hit-miss or misclassification indicator $\tau_{ijl} \in \{0,1\}$, where $0$ indicates that $\boldsymbol{y}_{ijl} = \boldsymbol{z}_{ijl}$. To model the observed classifications, we assume 
\begin{equation}\label{observedclass}
\boldsymbol{y}_{ijl}|\boldsymbol{\theta}_{t},\tau_{ijl}, \boldsymbol{z}_{ijl} \sim \tau_{ijl}\mbox{Multinomial}(1,\boldsymbol{\theta}_t) + (1-\tau_{ijl})\delta_{\boldsymbol{z}_{ijl}}(\boldsymbol{y}_{ijl}),
\end{equation}

\noindent where $\boldsymbol{\theta}_t$ is a $C$-dimensional vector of observed classification probabilities for the ${t}$th true classification and $\delta_{w}(\cdot)$ is a Dirac delta function at $w$. With this formulation, we assume that if there is no misclassification  (i.e., $\tau_{ijl} = 0$), then $\boldsymbol{y}_{ijl}= \boldsymbol{z}_{ijl}$, otherwise,  the individual is considered misclassified with $\tau_{ijl} = 1$. Note that this approach allows for $\boldsymbol{y}_{ijl}$ to be misclassified into the correct category (i.e., $\boldsymbol{y}_{ijl} = \boldsymbol{z}_{ijl}$, but $\tau_{ijl} = 1$). As such, our modeling approach places a positive probability of a ``lucky guess'' to occur. In section \ref{sec::postserior}, we discuss how to restrict the model to prevent this from occurring if desired. 

We assume the classification probabilities depend on the true classification of each individual with $\boldsymbol{\theta}_t \sim \mbox{Dirichlet}(\boldsymbol{\nu}_t)$, where $\boldsymbol{\nu}_t$ is a $C$-dimensional vector of concentration hyperparameters. To allow the classification probabilities to depend on an observed set of covariates, $\nu_{tc}$ can be replaced with a log-linear regression model similar to \cite{wadsworth2017integrative}. Next, we let the latent misclassification indicators, 
\begin{equation}\label{misclassifcation}
    \tau_{ijl}| \boldsymbol{z}_{ijl}, \boldsymbol{\beta}_{\psi_t}, \boldsymbol{x}_{ijl} \sim \mbox{Bernoulli}( \psi_{ijl}),
\end{equation}

\noindent where $\text{logit}( \psi_{ijl} ) =  \boldsymbol{x}_{ijl}^{\prime}\boldsymbol{\beta}_{\psi_t}$,  $\boldsymbol{x}_{ijl}$ is a $P_{\psi}$-dimensional vector of observed covariates that are observation-, measurement-, and/or individual-specific (including an intercept term), and $\boldsymbol{\beta}_{\psi_t}$ are the corresponding regression coefficients. We assume $\beta_{\psi_{tp}} \sim \mbox{Normal}(\mu_{\psi},\sigma_{\psi}^2)$. Note that the covariate effects on misclassification are allowed to vary based on the true classification of the individual. 

We model the true classification of each  individual 
\begin{equation}\label{trueclassification}
\boldsymbol{z}_{ijl}|\boldsymbol{\Theta}_{ij} \sim \text{Multinomial}(1,\boldsymbol{\Theta}_{ij}), 
\end{equation}
\noindent where $\boldsymbol{\Theta}_{ij}$ is a $T$-dimensional vector of true classification probabilities (or relative abundances), which we assume follows a Dirichlet$(\boldsymbol{\gamma}_{ij})$ with $\boldsymbol{\gamma}_{ij}$ a $T$-dimensional vector of concentration hyperparameters. With the availability of validation data, some of the $\boldsymbol{z}_{ijl}$ will be known and fixed in the model to inform the estimation of the classification matrix $\boldsymbol{\theta} = (\boldsymbol{\theta}_1^{\prime}, \dots,\boldsymbol{\theta}_T^{\prime})^{\prime}$, similar to \cite{wright2020modelling} and \cite{spiers2022estimating}.  More formally, the model can accommodate validation data by assuming $\boldsymbol{z}_{ijl}|\boldsymbol{\Theta}_{ij},v_{ijl} \sim (1- v_{ijl})\text{Multinomial} (1,\boldsymbol{\Theta}_{ij}) + v_{ijl}\delta_{\boldsymbol{y}_{ijl}}(\boldsymbol{z}_{ijl})$, where $v_{ijl} \in \{0,1\}$ is a fixed validation indicator for each individual. In practice, our model can accommodate any amount of validation data available, which, if error-free, can improve the estimation performance. By modeling the validation data at the individual level, the number of true categories does not have to equal the number of observed categories, which allows some categories to go unobserved, the potential for new categories to be observed, and the incorporation of individual-level covariates which can improve estimation, similar to \cite{spiers2022estimating}. We refer to $\boldsymbol{\theta}$ as a classification matrix to differentiate it from methods that use a ``confusion matrix,'' denoted as $\boldsymbol{\theta}^*$, to model false positives in multinomial data which, unlike our model, do not explicitly model misclassfication.  

We can equivalently model $\boldsymbol{\Theta}_{ij}$ as a set of independent gamma random variables normalized by their sum (i.e., $\boldsymbol{z}_{ijl}|\boldsymbol{\alpha}_{ij} \sim \text{Multinomial} (1,\frac{\alpha_{ijt}}{\bar{\alpha}_{ij}})$, where $\alpha_{ijt} \sim \text{Gamma}(\gamma_{ijt} , 1 )$ and $\bar{\alpha}_{ij} =  \sum_{t=1}^T \alpha_{ijt}$).  This reparameterization enables us to differentiate between at-risk or structural zeros (i.e., account for potential zero-inflation, false negatives, or non-detection) by introducing a latent at-risk (or occupancy) indicator $\zeta_{ijt}$ for the $t${th} category (or species\textcolor{black}{/crime}) at the $i${th} observation (or site\textcolor{black}{/jurisdiction}) and $j${th} measurement (or visit\textcolor{black}{/survey}).  Specifically, we instead let
\begin{equation}
   \alpha_{ijt}|\zeta_{ijt}, \gamma_{ijt} \sim \zeta_{ijt} \mbox{Gamma}(\gamma_{ijt},1) + (1- \zeta_{ijt})\delta_0(\alpha_{ijt}), 
\end{equation}
 similar to \cite{koslovsky2023bayesian}. \textcolor{black}{To model the relation between a set of observed covariates and the} latent at-risk (or occupancy) indicators, we assume $\zeta_{ijt}|\boldsymbol{\beta}_{\eta_{t}}, \boldsymbol{x}_{i} \sim \mbox{Bernoulli}(\eta_{it})$, where $\text{logit}(\eta_{it}) = \boldsymbol{x}_{i}^{\prime}\boldsymbol{\beta}_{\eta_{t}}$, $\boldsymbol{x}_{i}$ is a $P_{\eta}$-dimensional set of observation-specific covariates (including an intercept term), and $\boldsymbol{\beta}_{\eta_{t}}$ represent the corresponding true classification-specific regression coefficients. 
  We then let $ \beta_{\eta_{tp}} \sim \mbox{Normal}(\mu_{\eta},\sigma^2_{\eta})$. To allow the true classification probabilities (or relative abundances) to depend on a set of covariates, we set $\log(\gamma_{ijt}) =   \boldsymbol{x}_{ij}^{\prime} \boldsymbol{\beta}_{\gamma_{t}}$ with $\beta_{\gamma_{tp}} \sim \mbox{Normal}(\mu_{\gamma},\sigma^2_{\gamma})$ and $\boldsymbol{x}_{ij}$  an observation- and/or measurement-specific set of covariates.  In general, the model \textcolor{black}{does not require covariate information to inform the probability of an at-risk observation, the true classification probabilities, or the misclassification probabilities. In these settings, $\beta_{\eta_{t1}}$,  $\beta_{\gamma_{t1}}$, and/or $\beta_{\psi_{t1}}$  can be treated as hyperparameters that reflect prior beliefs for the corresponding probabilities. Likewise, it is straightforward to  adjust the model to accommodate observation-, measurement-, and/or, individual-level covariate information for the classification matrix concentration parameters, if available.}

\section{Posterior Sampling and Inference}\label{sec::postserior}

For posterior inference, we construct a Metropolis-Hastings within Gibbs sampler.  The full joint  distribution is defined as  
 
\begin{equation} 
\label{eq::postioer}
\begin{split}
& \prod_{i=1}^{N}\prod_{j=1}^{n_i} \prod_{l=1}^{L_{ij}} p(\boldsymbol{y}_{ijl}|\boldsymbol{\theta}_t,\tau_{ijl},\boldsymbol{z}_{ijl})  p(\boldsymbol{z}_{ijl}| \boldsymbol{\Theta}_{ij}) p(\tau_{ijl}|\boldsymbol{z}_{ijl},\boldsymbol{\beta}_{\psi_t},\boldsymbol{x}_{ijl})p(\omega_{\tau_{ijl}}) \\
& \times \prod_{i=1}^{N}    \prod_{j=1}^{n_i}  \prod_{t = 1}^{T}  p(\alpha_{ijt}|\zeta_{ijt},\boldsymbol{\beta}_{\gamma_t}, \boldsymbol{x}_{ij})p(\zeta_{ijt}|\boldsymbol{\beta}_{\eta_{t}}, \boldsymbol{x}_{i})p(\omega_{\zeta_{tij}} )    \\
& \times \prod_{i=1}^{N} \prod_{j=1}^{n_i} p(\mu_{ij}|\bar{\alpha}_{ij}) \prod_{t=1}^{T} \left[ p(\boldsymbol{\beta}_{\eta_{t}})p(\boldsymbol{\beta}_{\psi_t})p(\boldsymbol{\beta}_{\gamma_{t}})p(u_t|\bar{a}_t)  \prod_{c=1}^{C}  p(a_{tc})   \right],
\end{split}
\end{equation} 
where we introduce an auxiliary parameter $\mu_{ij}|\bar{\alpha}_{ij} \sim \mbox{Gamma}(1, \bar{\alpha}_{ij})$ for efficient sampling of $\alpha_{ijt}$.  The reparameterization of $\boldsymbol{\Theta}_{ij}$ with $\boldsymbol{\alpha}_{ij}$, coupled with the inclusion of $\mu_{ij}$, reduces the computational demand of updating $\boldsymbol{\beta}_{\gamma_t}$ and provides closed-form Gibbs updates for $\mu_{ij}$ and $\alpha_{ijt}|\zeta_{ijt} = 1$, which  greatly improves the overall scalability of the model to large $T$ settings. See \cite{koslovsky2020microbiome} and \cite{koslovsky2023bayesian} for more technical details of this data augmentation technique, its performance in high-dimensional settings, and parameter identifiability in this portion of the model. Similarly, we reparameterize $\theta_{tc} = a_{tc}/\bar{a}_t$ and assume $a_{tc} \sim \mbox{Gamma}(\nu_{tc}, 1 )$ with auxiliary parameter $u_t \sim \mbox{Gamma}( 1, \bar{a}_t )$ and $\bar{a}_t = \sum_{c=1}^C a_{tc}$. In addition to enabling efficient sampling of $\boldsymbol{\theta}_t$, this step provides the opportunity to easily incorporate covariates and restrict the model to disallow correct classifications by chance by fixing $a_{tt} = 0$.  Further, we exploit a P\'{o}lya-Gamma (PG) augmentation scheme that provides closed-form Gibbs updates for the regression coefficients associated with the latent at-risk (or occupancy) indicators, $\zeta_{ijt}$, and the misclassification indicators, $\tau_{ijl}$, without sacrificing their interpretability as log odds ratios following \cite{polson2013bayesian}. Specifically by introducing a latent set of auxiliary parameters $\omega_{\zeta_{tij}} \sim \mbox{PG}(1,0)$ and $\omega_{\tau_{ijl}} \sim \mbox{PG}(1,0)$, the full-conditional distributions of $\boldsymbol{\beta}_{\eta_t}$ and $\boldsymbol{\beta}_{\psi_t}$ are multivariate normal. A graphical representation of the proposed approach for modeling misclassification in zero-inflated Dirichlet-multinomial models, missZIDM, is presented in Figure \ref{dag}. More details of the Markov chain Monte Carlo (MCMC) sampler used to implement our model are provided in the Supplementary Material.

 Under the assumptions of the proposed model, $\beta_{\eta_{tp}}$ is interpreted as the expected change in log odds ratio of the $t${th} category  being at-risk at the $ij${th} measurement, and  $\exp(\beta_{\gamma_{tp}})$ is interpreted as the multiplicative change in the concentration parameter $\gamma_t$ for a one unit increase in the $p${th} covariate holding all else constant. While the latter provides inference about how concentrated the true classification probabilities (or relative abundances) are around the mean, we also are interested in inferring the relation between a covariate and the true classification probabilities directly. This inference is more complicated because each covariate is potentially associated with each category, as described in \cite{dai2019batch}. The multiplicative effect on the $t${th} category for a one unit increase in the $p${th} covariate for the $i${th} observation at the $j${th} measurement is defined as 
\begin{equation} \label{eq:pi}
  \pi_{ijtp} = \frac{\Theta_{ijt}(\boldsymbol{x}^{(p)}_{ij})}{\Theta_{ijt}( \boldsymbol{x}_{ij})} = \exp(\beta_{\gamma_{tp}})\frac{\sum_{s = 1}^T\exp(\boldsymbol{x}^{\prime}_{ij}\boldsymbol{\beta}_{\gamma_{s}})}{\sum_{s = 1}^T\exp(\boldsymbol{x}^{(p) \prime}_{ij} \boldsymbol{\beta}_{\gamma_{s}})},  
\end{equation}  
  \noindent where $\boldsymbol{x}^{(p) }_{ij} = (x_{ij1}, x_{ij2},\dots, x_{ijp} + 1, \dots, x_{ijP})^{\prime}$. Because the effect of the $p${th} covariate on the $t${th} category depends on its corresponding regression coefficient, $\beta_{\gamma_{tp}}$, in addition to its effect on the other categories and their corresponding concentration parameters, we may observe a decrease (increase) in the probability of the $t${th} category with an increase in $x_{ijp}$, even if $\beta_{\gamma_{tp}} > 0$ ($\beta_{\gamma_{tp}} < 0$). Estimates of the true classification probabilities for each observation can be obtained by normalizing the vector $\boldsymbol{\alpha}_i$ over its sum for each MCMC iteration and then averaging over the samples. Estimates of the classification matrix $\boldsymbol{\theta}$ can be obtained similarly given $\boldsymbol{a}_{t}$. While we model the probability of misclassification separately from the classification matrix, we can generate inference on the confusion matrix, $\boldsymbol{\theta}^*$, typically estimated by other approaches. Assuming an intercept-term only model for the probability of misclassification for simplicity, each element of $\boldsymbol{\theta}^*$, $\theta^*_{tc} = \theta_{tc}\frac{\exp(\beta_{\psi_{t1}})}{1 + \exp(\beta_{\psi_{t1}})} +  [ 1 - \frac{\exp(\beta_{\psi_{t1}})}{1 + \exp(\beta_{\psi_{t1}})} ]I( t = c),$
where $I(\cdot)$ is an indicator function. For inference on these quantities, the posterior means of the MCMC samples are calculated and 95\% credible intervals are constructed using the empirical quantiles.

\begin{figure}[ht] 
\centering
    \begin{tikzpicture}[x=1cm,y=1cm]
        \node[latent, scale = 1.2] (zijl) {$\boldsymbol{z}_{ijl}$} ; %
        \node[obs, above = of zijl, scale = 1.2] (yijl) {$\boldsymbol{y}_{ijl}$} ; %
        \node[latent, below = of zijl, scale = 1.2, yshift = 0.2cm] (tauijl) {$\tau_{ijl}$} ; 
        \node[latent, left= of yijl, scale = 1.2, xshift = -2.2cm, yshift = 1.7cm] (thetatc) {$\boldsymbol{\theta}_{tc}$} ; 
        \node[latent, left= of zijl, yshift = -0cm, xshift = -0.75cm, scale = 1.2] (alphaijt) {$\alpha_{ijt}$} ; 
         \node[rectangle,draw, fill = black!10, right= of alphaijt, xshift = -.5cm, yshift = -.75cm, scale = 1.2] (xij) {$\boldsymbol{x}_{ij}$} ; 
        \node[latent, left= of zijl, yshift = 1.3cm, xshift = -0.75cm, scale = 1.2] (zetaijt) {$\zeta_{ijt}$} ; 
        \node[latent, draw, left= of alphaijt,  yshift = -0cm, xshift = 0.25cm, scale = 1.2] (gammaijt) {$\boldsymbol{\beta}_{\gamma_{t}}$} ; 
        
        \node[rectangle, draw, fill = black!10, above= of zetaijt, xshift = 1.2cm, yshift = -0cm, scale = 1.2] (xi) {$\boldsymbol{x}_{i}$} ; %
        
       \node[latent, draw, left= of zetaijt, xshift = 0.25cm,  yshift= 0cm, scale = 1.2] (etait) {$\boldsymbol{\beta}_{\eta_{t}}$} ; %

         \node[latent, left = of tauijl, xshift = -2.35cm,   scale = 1.2] (betapsit) {$\boldsymbol{\beta}_{\psi_{t}}$} ; %

         \node[rectangle,draw, fill = black!10, below= of tauijl, yshift = 0.5cm, scale = 1.2] (xijl) {$\boldsymbol{x}_{ijl}$} ; 
         
         \node[ below = of tauijl , xshift = -3.4cm, yshift = -.4cm]  (N) { $N$ };
         \node[ below = of tauijl , xshift = -3cm, yshift = -.15cm]  (ni) { $n_i$ };
        \node[ below = of tauijl , xshift = 0.35cm, yshift = -.0cm]  (Lij) { $L_{ij}$ };
        \node[ below = of thetatc , xshift = -0.5cm, yshift = -4.6cm]  (T) { $T$ };
        \node[ below = of thetatc , xshift = .35cm, yshift =  1.15cm]  (C) { $C$ };
   
     \path[->] (tauijl) edge[bend right = 50 ] node [left] {} (yijl); 
      \path[->] (thetatc) edge[bend left = 50 ] node [left] {} (yijl); 
    \edge {zijl} {yijl} ; %
    \edge {zijl} {tauijl} ;
    \edge {alphaijt} {zijl} ;
    \edge {zetaijt}  {zijl} ;
    \edge {gammaijt} {alphaijt}   ;
    \edge {etait} {zetaijt}   ;
    \edge {betapsit} {tauijl} ;
     \edge {xijl} {tauijl} ;
    \edge {xij} {alphaijt} ;
    \edge {xi} {zetaijt} ;

    \plate[inner sep=.5cm, color=black] {N} { (zijl) (yijl) (tauijl) (zetaijt) (alphaijt)   (xij) (xi) (xijl)   } {  }; 
  \plate[inner sep=.2cm, color=black] {ni} { (zijl) (yijl) (tauijl) (zetaijt) (alphaijt)   (xij) (xijl)    } {  }; 
    \plate[inner sep=.15cm, color=black] {Lij} { (zijl) (yijl) (tauijl)   (xijl)    } {  }; 
    \plate[inner sep=.3cm, color=black] {T} { (thetatc)  (zetaijt) (alphaijt)  (gammaijt) (betapsit) (etait)   } {  }; 
   \plate[inner sep=.1cm, color=black] {C} { (thetatc)    } {  }; 
\end{tikzpicture}
  \caption{Graphical representation of missZIDM with covariate dependence. Note that auxiliary parameters and hyperparameters have been suppressed for clarity. $\boldsymbol{N}$ - total observations; $\boldsymbol{n}_i$ - total measurements per observation; $\boldsymbol{L}_{ij}$ - number of individuals at each measurement; $\boldsymbol{T}$ - true number of categories; $\boldsymbol{C}$ - observed number of categories. }\label{dag}
\end{figure}

\section{Empirical Studies}
 In this section, we first compare the proposed model to alternative methods for handling false positives or false negatives in multinomial data \textcolor{black}{ in  simulated settings without covariate information, repeated measurements, or validation data to inform the model, similar to the official crime data application.} In a second scenario, we compare the proposed model to a multispecies occupancy model in settings designed to mimic the \textcolor{black}{bat monitoring} study.

The first scenario examines the estimation performance of missZIDM with respect to the at-risk probability, $\boldsymbol{\eta}$, the probability of misclassification, $\boldsymbol{\psi}$, the true probability of each category, $\boldsymbol{\Theta}$, and the confusion matrix, $\boldsymbol{\theta}^*$, at varying percentages of at-risk observations and misclassification.  In the first scenario, we compare missZIDM to a similar approach that does not accommodate false negatives (missDM), an approach that assumes the true and observed classifications follow Dirichlet-multinomial models which does not explicitly model misclassification or handle false negatives (DMDM, similar to \cite{swartz2004bayesian}), and the recently developed zero-inflated Dirichlet-multinomial model (ZIDM; \cite{koslovsky2023bayesian}), which is designed to accommodate false negatives but not false positives.  In this scenario, the models ignore any potential covariates and therefore only estimate intercept terms for $\boldsymbol{\eta}$, $\boldsymbol{\psi}$, and $\boldsymbol{\Theta}$. For the DMDM model, we set $\nu_{tt} = ( \mbox{logit}^{-1}(\mu_\psi)/T + ( 1- \mbox{logit}^{-1}(\mu_\psi)) )\times T/\mbox{logit}^{-1}(\mu_\psi)$, which places a similar prior probability for correct classification as the methods with misclassification indicators. All methods were implemented in \texttt{R} using \texttt{Rcpp} \citep{eddelbuettel2011rcpp}.

 In this scenario, we generated $N=50$ observations of $L_{ij} = 100$ individuals to cluster into $C=10$ categories. We assumed that the true number of categories, $T$, matched the potentially observed number of categories, $C$. We evaluated the model in four settings with varying percentages of at-risk observations (1 - \% true negatives) and misclassification (false positives). In these settings, we set $n_i = 1$ (i.e., no repeated measurements). Observation-specific at-risk indicators were sampled from a Bernoulli distribution with the probability of an at-risk observation set to either 0.25 or 0.75. The true classification of each individual was generated from a Dirichlet-multinomial distribution with concentration parameters set to one and overdispersion parameter set to 0.01, so that the model assumptions did not match the true data generation process. Misclassification occurred with 0.25 or 0.75 probability. The observed classifications were generated from a Dirichlet-multinomial model with a similar overdispersion parameter as above. We set concentration parameters $\boldsymbol{\nu}_{t}$ equal to their index (e.g., $\nu_{tc} = c$ ) with the $t$th element also equal to one, placing the least probability on a correct classification by chance. No validation data were used in these settings.
 
 In the second scenario, we investigate how the proposed method performs when used for inference in multispecies occupancy-detection settings with data generated to mimic the \textcolor{black}{bat monitoring} data set. We evaluate the performance of the method with varying percentages of overdispersion in the true counts.    We compare the proposed model to a similar version of the multispecies occupancy-detection model presented in \cite{wright2020modelling}, which we refer to as DMZIP. This approach differs from the proposed model in that it makes distributional assumptions for the ecological process, does not explicitly model the latent classifications, and uses a confusion matrix to  accommodate potential misclassification. Using the notation of our proposed model, DMZIP assumes the detection counts of each species $\sum_{l=1}^{L_{ij}}I(z_{ijlt} =1 ) \sim \mbox{Poisson}(\lambda_{ijt}\zeta_{ijt})$, where $\zeta_{ijt} \sim \mbox{Bernoulli}(\eta_{it})$. Given the true latent cluster, the observed individual classifications $\boldsymbol{y}_{ijl} \sim \mbox{Multinomial}(\boldsymbol{\theta}^*_t)$, where $\boldsymbol{\theta}^*_t \sim \mbox{Dirichlet}(\boldsymbol{\gamma}_{ij})$. Code to implement DMZIP was adapted from \cite{Stratton2022}. 

 Specifically, we generated $N=50$ sites with $n_i = 5$ visits per site and $C = 10$ possible species to observe. No covariates were used in the baseline simulation setting.  Parameter values for the occupancy probabilities, encounter rates, and confusion matrix used to simulate the data were obtained from the posterior mean estimates obtained with the count detection model fit to acoustic data in \cite{stratton2022coupling}. The true occurrence probabilities ranged from 23\% to 90\%. Instead of assuming the total number of counts was fixed, as in scenario 1, the relative activity or encounter rates for each species at each site visit was generated from a negative binomial distribution with mean $\zeta_{it} * \lambda_{t} $ and variance $(\zeta_{it} * \lambda_{t})^2/\sigma$, where $\zeta_{it}$ is the site-level occupancy indicator for a given species, $\lambda_{t}$ is the expected number of detections or encounter rate of species $t$ obtained from the \textcolor{black}{bat monitoring} data which ranged from 2.0 to 28.2, and $\sigma$ is an overdispersion parameter. Note that as $\sigma$ increases, the variance of the sampling distribution approaches the mean, and the data are more Poisson-like. The models were evaluated with $\sigma \in \{0.1,1,100\}$  and 25\% validation data. 

  We then evaluated model performance on data generated similar to scenario 2 but with covariates informing the occupancy probability and the true encounter rates. We simulated 5 continuous covariates from a standard normal distribution in both levels of the model. In this setting, we set the overdispersion parameter for the negative binomial distribution $\sigma = 1$. The intercept terms $\beta_{\eta_{t0}}$ ($\beta_{\gamma_{t0}}$) were randomly sampled uniformly from $\mbox{logit}(0.25)$  to  $\mbox{logit}(0.95)$  ($0$ to $\log(10)$) with covariate effects set to $\pm 1$ ($\pm 0.2)$ with equal probability. The off-diagonal elements of the classification matrix $\boldsymbol{\theta}$ were sampled uniformly from 0.01 to 0.2 with diagonal elements uniformly sampled from 0.5 to 0.95. Thereafter, the rows of $\boldsymbol{\theta}$ were scaled to sum to one, and the individual classifications were sampled from a Multinomial$(1, \boldsymbol{\theta}_t)$. In this setting, we assumed 25\% of the data were validated. Additionally, we evaluated the models in various other data generation settings including those with  different sample sizes, sampling efforts, and percent validated data.

  In both scenarios, each of the MCMC algorithms were run for 5,000 iterations treating the first 2,500 as burn-in and thinning to every other iteration, providing 1,250 iterations for inference. We assumed non- or weakly-informative priors $\gamma_{tc} = \nu_{tc} = \sigma_{\eta}^2= \sigma_{\psi}^2= \sigma_{\gamma}^2  = 1$. In settings with no covariates in the model, we set $\mu_\eta$ and $\mu_\psi$ following the data generation process for all models. In settings with covariates in the model, the prior mean for the regression coefficients was set to 0. In the sensitivity analysis presented in the Supplementary Material, we explore the impact of prior misspecification of these hyperparameters on inference. To initialize each model, we set the true classifications $\boldsymbol{Z}_i$ to the observed classifications $\boldsymbol{Y}_i$, with $\boldsymbol{\tau}_i$ set accordingly. Auxiliary parameters, $\omega_{\tau_{tijl}}$ and $\omega_{\zeta_{tij}}$, and at-risk indicators, $\zeta_{tij}$ and $\alpha_{tc}$, were initialized at one.  The auxiliary parameters $u_t$ and $\mu_{ij}$ were randomly initialized from a Gamma(1,1) and regression coefficients were set to 0. 
  
  We evaluated the models in terms of the average absolute value of the difference between the estimated and true probabilities (ABS) and Frobenius norm (FROB), which is the square root of the sum of the squared difference of the estimated and true probabilities for $\boldsymbol{\eta}$, $\boldsymbol{\psi}$, $\boldsymbol{\theta}^*$, and $\boldsymbol{\Theta}$.  Note that the proposed method is the only method that provides estimates for all parameters simultaneously. In settings where covariates were incorporated into the data generation process (results presented in the Supplementary Material), the models were compared with respect to the estimation of the occupancy probability regression coefficients, $\boldsymbol{\beta}_\eta$, and the confusion matrix, $\boldsymbol{\theta}^*$, because these maintained similar interpretation among all models.  
  Results we report below were obtained by averaging over 50 replicated data sets for each setting. 

\subsection{Results}

   In the first scenario, the estimation performance for the probability of an at-risk observation, $\boldsymbol{\eta}$, improved as the true percentage of at-risk observations increased for missZIDM and ZIDM, with missZIDM demonstrating better performance than ZIDM in settings with more structural zeros (i.e., 25\% at-risk observations) regardless of the amount of misclassification (Table \ref{tab::scenario1}).  We observed that the proposed missZIDM always outperformed missDM, which ignores potential zero-inflation, when estimating the probability of misclassification, $\boldsymbol{\psi}$. All methods obtained relatively similar estimation performance for the true classification probabilities, $\boldsymbol{\Theta}$, with the proposed method demonstrating a slight advantage in the setting with 25\% at-risk observations and misclassification. Estimation accuracy for the confusion matrix, $\boldsymbol{\theta}^*$, reduced as the percentage of misclassification increased for all methods. The proposed method and DMDM both outperformed missDM with respect to estimating the confusion matrix, $\boldsymbol{\theta}^*$. However, DMDM obtained a two-fold reduction in the absolute value of the bias compared to missZIDM when data were generated with greater misclassification percentages (0.04 and 0.08, respectively). Recall that missZIDM, unlike DMDM, does not directly provide estimates for $\boldsymbol{\theta}^*$, but they can be obtained using the estimated $\boldsymbol{\psi}$ and $\boldsymbol{\theta}$ values. 

 \begin{table}[ht]
\caption{ Simulation Results for Scenario 1: Estimation performance for $N=50$ observations, $n_i = 1$ measurements, $L_{ij} = 100$ individuals, and $T=10$ categories at varying percentages of at-risk observations and misclassification with 0\% validation data.  Bold indicates the best performing models. Standard deviations of performance metrics across the replicate data sets are provided in parentheses. ABS - absolute value of the difference between the estimated and true probabilities; FROB - Frobenius norm. } \label{tab::scenario1}  
\setlength{\tabcolsep}{1pt} 
\scriptsize
\centering
\begin{tabular}{cccccccccccc}
\hline
         & \multicolumn{11}{c}{25\% at-risk observations  and 25\% misclassification}                                                                                                                              \\ \hline
         & \multicolumn{2}{c}{$\eta$}                  &  & \multicolumn{2}{c}{$\psi$}                  &           & \multicolumn{2}{c}{$\Theta$}                &  & \multicolumn{2}{c}{$\theta^*$}              \\ \cline{2-3} \cline{5-6} \cline{8-9} \cline{11-12} 
         & ABS                  & FROB                 &  & ABS                  & FROB                 &           & ABS                  & FROB                 &  & ABS                  & FROB                 \\ \cline{2-3} \cline{5-6} \cline{8-9} \cline{11-12} 
missZIDM & \textbf{0.05 (0.02)} & \textbf{0.25 (0.05)} &  & \textbf{0.02 (0.01)} & \textbf{0.07 (0.02)} &           & \textbf{0.02 (0.00)} & \textbf{0.71 (0.06)} &  & \textbf{0.01 (0.00)} & \textbf{0.14 (0.02)} \\
missDM   & -                    & -                    &  & 0.18 (0.01)          & 0.56 (0.02)          &           & 0.05 (0.00)          & 1.49 (0.07)          &  & 0.04 (0.00)          & 0.62 (0.02)          \\
DMDM     & -                    & -                    &  & -                    & -                    &           & 0.05 (0.00)          & 1.66 (0.08)          &  & 0.01 (0.00)          & 0.15 (0.00)          \\
ZIDM     & 0.13 (0.01)          & 0.43 (0.02)          &  & -                    & -                    &           & 0.05 (0.00)          & 1.63 (0.08)          &  & -                    & -                    \\ \hline
         & \multicolumn{11}{c}{25\% at-risk observations and 75\% misclassification}                                                                                                                               \\ \cline{2-12} 
         & \multicolumn{2}{c}{$\eta$}                  &  & \multicolumn{2}{c}{$\psi$}                  &           & \multicolumn{2}{c}{$\Theta$}                &  & \multicolumn{2}{c}{$\theta^*$}              \\ \cline{2-3} \cline{5-6} \cline{8-9} \cline{11-12} 
         & ABS                  & FROB                 &  & ABS                  & FROB                 &           & ABS                  & FROB                 &  & ABS                  & FROB                 \\ \cline{2-3} \cline{5-6} \cline{8-9} \cline{11-12} 
missZIDM & \textbf{0.09 (0.01)} & \textbf{0.33 (0.03)} &  & \textbf{0.36 (0.01)} & \textbf{1.15 (0.04)} & \textbf{} & \textbf{0.11 (0.00)} & 3.76 (0.14)          &  & 0.08 (0.00)          & 1.29 (0.04)          \\
missDM   & -                    & -                    &  & 0.47 (0.01)          & 1.49 (0.03)          &           & 0.12 (0.00)          & \textbf{3.75 (0.16)} &  & 0.09 (0.00)          & 1.61 (0.03)          \\
DMDM     & -                    & -                    &  & -                    & -                    &           & 0.12 (0.00)          & 3.93 (0.18)          &  & \textbf{0.04 (0.00)} & \textbf{0.45 (0.00)} \\
ZIDM     & 0.14 (0.01)          & 0.45 (0.01)          &  & -                    & -                    &           & 0.12 (0.00)          & 3.93 (0.18)          &  & -                    & -                    \\ \hline
         & \multicolumn{11}{c}{75\% at-risk observations and 25\% misclassification}                                                                                                                               \\ \cline{2-12} 
         & \multicolumn{2}{c}{$\eta$}                  &  & \multicolumn{2}{c}{$\psi$}                  &           & \multicolumn{2}{c}{$\Theta$}                &  & \multicolumn{2}{c}{$\theta^*$}              \\ \cline{2-3} \cline{5-6} \cline{8-9} \cline{11-12} 
         & ABS                  & FROB                 &  & ABS                  & FROB                 &           & ABS                  & FROB                 &  & ABS                  & FROB                 \\ \cline{2-3} \cline{5-6} \cline{8-9} \cline{11-12} 
missZIDM & 0.03 (0.01)          & 0.12 (0.04)          &  & \textbf{0.03 (0.01)} & \textbf{0.10 (0.04)} &           & 0.03 (0.00)          & 0.84 (0.03)          &  & 0.01 (0.00)          & 0.17 (0.03)          \\
missDM   & -                    & -                    &  & 0.09 (0.02)          & 0.28 (0.05)          &           & 0.03 (0.00)          & 0.82 (0.02)          &  & 0.02 (0.00)          & 0.35 (0.05)          \\
DMDM     & -                    & -                    &  & -                    & -                    &           & 0.03 (0.00)          & 0.81 (0.02)          &  & \textbf{0.01 (0.00)} & \textbf{0.15 (0.00)} \\
ZIDM     & \textbf{0.02 (0.01)} & \textbf{0.07 (0.04)} &  & -                    & -                    &           & \textbf{0.03 (0.00)} & \textbf{0.81 (0.02)} &  & -                    & -                    \\ \hline
         & \multicolumn{11}{c}{75\% at-risk observations and 75\% misclassification}                                                                                                                               \\ \cline{2-12} 
         & \multicolumn{2}{c}{$\eta$}                  &  & \multicolumn{2}{c}{$\psi$}                  &           & \multicolumn{2}{c}{$\Theta$}                &  & \multicolumn{2}{c}{$\theta^*$}              \\ \cline{2-3} \cline{5-6} \cline{8-9} \cline{11-12} 
         & ABS                  & FROB                 &  & ABS                  & FROB                 &           & ABS                  & FROB                 &  & ABS                  & FROB                 \\ \cline{2-3} \cline{5-6} \cline{8-9} \cline{11-12} 
missZIDM & 0.03 (0.01)          & 0.11 (0.03)          &  & \textbf{0.40 (0.01)} & \textbf{1.26 (0.03)} &           & 0.07 (0.00)          & 1.99 (0.08)          &  & 0.08 (0.00)          & 1.40 (0.03)          \\
missDM   & -                    & -                    &  & 0.44 (0.01)          & 1.40 (0.02)          &           & 0.06 (0.00)          & 1.70 (0.07)          &  & 0.09 (0.00)          & 1.53 (0.02)          \\
DMDM     & -                    & -                    &  & -                    & -                    &           & 0.06 (0.00)          & 1.53 (0.07)          &  & \textbf{0.04 (0.00)} & \textbf{0.45 (0.00)} \\
ZIDM     & \textbf{0.01 (0.00)} & \textbf{0.04 (0.01)} &  & -                    & -                    &           & \textbf{0.06 (0.00)} & \textbf{1.53 (0.07)} &  & -                    & -                    \\ \hline
\end{tabular}
\end{table}

   In scenario 2, we found the DMZIP model obtained the best estimation performance for $\boldsymbol{\beta}_{\eta}$ and $\boldsymbol{\theta}^*$ when $\sigma = 100$ (Table \ref{tab::scenario2.2}). However in settings with more overdispersion ($\sigma = 1$ and  $0.1$), the proposed method  outperformed DMZIP with respect to both parameters. Notably, estimation performance was  worse for both methods when $\sigma = 0.1$. Similar trends were observed with 75\% validated data (Supplementary Table S1). These results demonstrate how the proposed method is preferred in the presence of overdispersion.  Additionally, we found that missZIDM obtained the best estimation performance for $\boldsymbol{\beta}_\eta$ and $\boldsymbol{\theta}^*$ with covariates in the model, more species types, visits, and sites (Supplementary Table S2).

 \begin{table}[ht]
\caption{  Simulation Results for Scenario 2: Estimation performance for data generated similar to the \textcolor{black}{bat monitoring data} with 25\% validation.  Bold indicates the best performing models. Standard deviations of performance metrics across the replicate data sets are provided in parentheses. ABS - absolute value of the difference between the estimated and true probabilities; FROB - Frobenius norm.\label{tab::scenario2.2}}
\setlength{\tabcolsep}{1pt} 
\centering
\scriptsize
\begin{tabular}{cccccccccccccccccccccccccccccc} 
\hline
  & \multicolumn{6}{c}{$\sigma = 0.1$} \\ 
\hline
   & \multicolumn{2}{c}{$\boldsymbol{\eta}$}        &     & \multicolumn{2}{c}{$\boldsymbol{\theta}^*$}                         \\ 
 \cline{2-3} \cline{5-6}  
  & \multicolumn{1}{c}{ABS} & \multicolumn{1}{c}{FROB} &   & \multicolumn{1}{c}{ABS} & \multicolumn{1}{c}{FROB}   \\ \cline{2-3} \cline{5-6}  
missZIDM  & \textbf{0.27 (0.07)} & \textbf{0.94 (0.20)}  & & \textbf{0.01 (0.00)} & \textbf{0.25 (0.01)}    \\  
DMZIP & 0.46 (0.01) & 1.54 (0.02)   &  & 0.04 (0.01) & 1.24 (0.02)   \\ 
\hline
  & \multicolumn{6}{c}{$\sigma = 1$} \\ 
\cline{2-6}
  & \multicolumn{2}{c}{$\boldsymbol{\eta}$}        &     & \multicolumn{2}{c}{$\boldsymbol{\theta}^*$}                       \\ 
\cline{2-3} \cline{5-6}  
  & \multicolumn{1}{c}{ABS} & \multicolumn{1}{c}{FROB} &   & \multicolumn{1}{c}{ABS} & \multicolumn{1}{c}{FROB} &     \\ \cline{2-3} \cline{5-6}  
missZIDM  & \textbf{0.15 (0.02)} & \textbf{0.60 (0.08)} &  & \textbf{0.01 (0.00)} & \textbf{0.22 (0.02)}   \\  
DMZIP & 0.22 (0.01) & 0.76 (0.04) &    & 0.03 (0.00) & 1.04 (0.05    \\ 
\hline 
   & \multicolumn{6}{c}{$\sigma = 100$} \\ 
\cline{2-6}
  & \multicolumn{2}{c}{$\boldsymbol{\eta}$}        &     & \multicolumn{2}{c}{$\boldsymbol{\theta}^*$}                      \\ 
  \cline{2-3} \cline{5-6}  
  & \multicolumn{1}{c}{ABS} & \multicolumn{1}{c}{FROB} &    & \multicolumn{1}{c}{ABS} & \multicolumn{1}{c}{FROB}    \\ \cline{2-3} \cline{5-6}   
missZIDM  & 0.17 (0.01) & 0.66 (0.06)  & & 0.02 (0.00) & 0.28 (0.03)    \\  
DMZIP & \textbf{0.03 (0.01)} & \textbf{0.12 (0.03)}   &  & \textbf{0.01 (0.00)} & \textbf{0.15 (0.03)}     \\ 
\hline  
\end{tabular}
\end{table}  

     One of the major challenges of modeling measurement error in multinomial data is non-identifiability of the parameters, because there is no information contained in the raw data to inform zero-inflation or misclassification probabilities. Our approach is designed to account for non-identifiability through informative prior specifications and/or incorporating a subset of validation data to inform parameter estimates. As such, inferential results obtained by our method, and any method designed to model measurement error in multinomial data, will be sensitive to the amount of validation data used to inform the model as well as the specification of the hyperparameters. In the Supplementary Material, we present an extensive sensitivity analysis of the proposed model with varying percentages of validated data (Supplementary Tables S3 and S4) and hyperparameter specification (Supplementary Tables S5 and S6). Based on these results, we found that validation data are useful when available. With only 10\% validation data, the proposed method was able to obtain less than 0.05\% bias for all probability estimates on average. In the absence of validation data, the model performed better with lower concentration parameters for the true classification probabilities, or relative abundances, and the observed classification probabilities. Additionally, we found that the model was relatively robust to misspecification of the misclassification prior for all other parameters. Similar results were observed for changes in the prior for the at-risk probability.

\section{\textcolor{black}{Real Data Applications}}

\textcolor{black}{ In this section, we apply the proposed method to \textcolor{black}{two publicly available data sets}. \textcolor{black}
{In section \ref{sec::bats}}, we show how the proposed method can serve as an
alternative approach for accommodating imperfect detection in multispecies occupancy-detection modeling. For this analysis, we incorporate validation data to inform the true relative abundances, classification probabilities, and occupancy probabilities. \textcolor{black}{In section \ref{sec::crime}}, we demonstrate how to accommodate potential zero-inflation and misclassification in official crime data in an unsupervised setting.  }

\subsection{\textcolor{black}{Application to Multispecies Bat Acoustic Monitoring  Data}}\label{sec::bats}


    \textcolor{black}{The goal of occupancy modeling in ecological research is to draw inference on species' true occurrence given a set of observations that are subject to imperfect detection due to observational error. Imperfect detection typically occurs in two different ways: (1) a species may go undetected and (2) an observed individual may be misclassified. Even with increased sampling effort, imperfect detection may still occur, resulting in biased inference if ignored when modeling  \citep{kellner2014accounting}.}  

 \textcolor{black}{ Historically, statistical methods developed to handle imperfect detection have focused on non-detection \citep{hoeting2000improved,bayley2001approach,mackenzie2002estimating, royle2003estimating, mackenzie2003estimating, tyre2003improving, broms2015accounting, dorazio2006estimating,dorazio2011modern,devarajan2020multi}. However, more recently researchers have proposed methods that account for both non-detection and misclassification, in part due to the emergence of automated species detection methods (e.g., unmanned aerial systems and automated recording units) and volunteer-based surveys (e.g., citizen science) for monitoring wildlife populations \citep{mcclintock2010unmodeled}. When developing single- or multispecies (community) occupancy models, researchers typically take a hierarchical approach, often referred to as occupancy-detection models, which jointly model the ecological and observation (or detection) process.  This technique allows researchers to differentiate between latent  species occupancy and observed species detection and effectively account for potential misclassification. Typically, this is achieved by introducing a site- or location-specific latent species indicator that models whether or not a species is present, reminiscent of the latent at-risk indicator for handling zero-inflation in count data. If the species is present (absent) at that site, there is a positive (zero) probability of detecting it.  See \cite{blasco2019does} for an in-depth discussion of zero counts in the context of ecological research studies, \cite{scharf2022constructing} for an overview of hierarchical models for occupancy data, and \cite{mackenzie2017occupancy} for more background on methods for estimating and modeling occupancy.}


 \textcolor{black}{Methods that handle potential species misclassification in occupancy modeling were initially developed for single species studies 
\citep{royle2006generalized,miller2011improving,chambert2015modeling, ruiz2016uncertainty,chambert2018new}. \cite{chambert2018two} introduced a two-species occupancy model that accounts for both species misidentification and non-detection. Their approach is based on the premise that false detections for a given species occur due to the misidentification with a closely related species. Recently, \cite{wright2020modelling} developed a multispecies occupancy model that handles both forms of measurement error for two or more species at each site visit. 
By assuming (1) the true count of each species follows a Poisson distribution given it is present at the site visit, (2) the number of detections for each species follows a multinomial distribution given the true species counts, and (3) the detection counts are independent across species, the authors demonstrate how the observed/detected counts can be directly modeled without conditioning on the true count for each species. \cite{spiers2022estimating} developed a multispecies occupancy model similar to \cite{wright2020modelling} that accommodates individual-level validation data (as opposed to site-level) which allows for more flexibility when modeling heterogeneity with covariates and morphospecies.} 
 

    In this analysis, we demonstrate the proposed method on data collected in a multispecies bat acoustic monitoring study conducted in British Columbia, Canada between 2016 and 2020. Details of the study design and data are found in \cite{stratton2022coupling} and \cite{Stratton2022}.  Briefly, one to six stationary acoustic recording devices were placed in $N=55$ sites following the North American Bat Monitoring Program guidelines and were typically activated for seven nights \citep{loeb2015plan}. Similar to  \cite{stratton2022coupling}, we analyze detections from the first and last nights to minimize potential overlap and dependencies, leading to  $n_i=$  2 to 12 measurements for each site over the five year period. There were  $T = C = 10$ total bat species categories available for analysis, including an \textit{other} category for species that were difficult to detect acoustically or that were not widespread. Each acoustic recording was classified using Kaleidoscope Pro acoustic classification software for bats (https://www.wildlifeacoustics.com). A unique attribute of these data is that each acoustic recording was additionally validated by a bat expert. For this analysis, we let half of all revisits from every site include validation data, similar to \cite{stratton2022coupling}. Additionally, we included site-specific covariates,  $\boldsymbol{x}_i$, measuring year (categorical with 2016 as the reference), annual mean elevation (kilometers), precipitation (millimeters), and temperature (degrees Celsius) for the occupancy (or at-risk) portion of the model.  We included nightly minimum air temperature (degrees Celsius), total precipitation (millimeters), and percentage of the moon illuminated by the sun (percent) measured from the centroid of the site at each visit to model the encounter rates or relative abundances for the DMZIP and missZIDM models, respectively. 
    All continuous covariates were standardized prior to analysis.  In this analysis, the average number of observed individuals per site visit was 62.6, ranging from 1 to 992. The means, variances, and percent zero counts for each species across site visits are presented in Table \ref{tab::commonname}.  We observed mean counts ranging from 1.7 to 25.9 with variances ranging from 25.7 to 4339.6 and percent zeros ranging from 26\% to 83\%.

\begin{table}[ht]
\hspace{-1.5cm}
\begin{tabular}{cccc}
\hline
Common Name                 & Scientific Name                  & Mean (Variance) & \% Zero \\ \hline
Big brown bat               & \textit{Eptesicus fuscus} (EPFU)          & 3.5 (256.0)     & 66      \\
Hoary bat                   & \textit{Lasiurus cinereus} (LACI)         & 3.1 (396.8)     & 65      \\
Silver-haired bat           &\textit{Lasionycteris noctivagans} (LANO) & 12.4 (2703.4)   & 34      \\
California myotis           & \textit{Myotis californicus} (MYCA)       & 4.5 (264.3)     & 54      \\
Western small-footed myotis & \textit{Myotis ciliolabrum} (MYCI)        & 2.9 (265.4)     & 83      \\
Western long-eared myotis   & \textit{Myotis evotis} (MYEV)             & 1.7 (28.2)      & 63      \\
Little brown myotis         & \textit{Myotis lucifugus} (MYLU)          & 25.9 (4339.6)   & 26      \\
Long-legged myotis          & \textit{Myotis volans} (MYVO)             & 2.7 (114.3)     & 59      \\
Yuma myotis                 & \textit{Myotis yumanensis} (MYYU)        & 5.0 (625.5)     & 69      \\
Other                       & -                                & 0.98 (25.7)     & 77      \\ \hline
\end{tabular}
\caption{Observed mean, variance, and \% zero observations for each species in the \textcolor{black}{bat monitoring} study. \label{tab::commonname} }
\end{table}

 For inference, missZIDM was run for 10,000 iterations, treating the first 5,000 as burn-in and thinning to every other iteration. The model was initialized similar to scenario 2 of the simulation study. We assumed relatively weak or non-informative priors with  $\nu_{tc} = 1$, $\gamma_t = 1$, $\mu_{\beta_\eta}=\mu_{\beta_\psi} = 0$, and $\sigma_\eta^2 =  \sigma_\psi^2 =\sigma_\gamma^2 = 1$. We compared the results of the proposed model to DMZIP with similar prior assumptions. Convergence and mixing of the models was visually inspected using traceplots.  See Supplementary Material for traceplots for a random subset of the parameters (Supplementary Figures S1-S7). To further assess the convergence of the model, we ran another chain initialized with   $\omega_{\tau_{tijl}} = \omega_{\zeta_{tij}} =\zeta_{tij} =\alpha_{tc}=0.5$, $u_t$, $\mu_{ij} \sim$ Gamma(1,1), and regression coefficients sampled from a standard normal. We then compared the two chains with the Gelman-Rubin statistic, which was less than 1.1 for each parameter \citep{brooks1998general}.

\begin{figure}[ht]
\includegraphics[width=\textwidth]{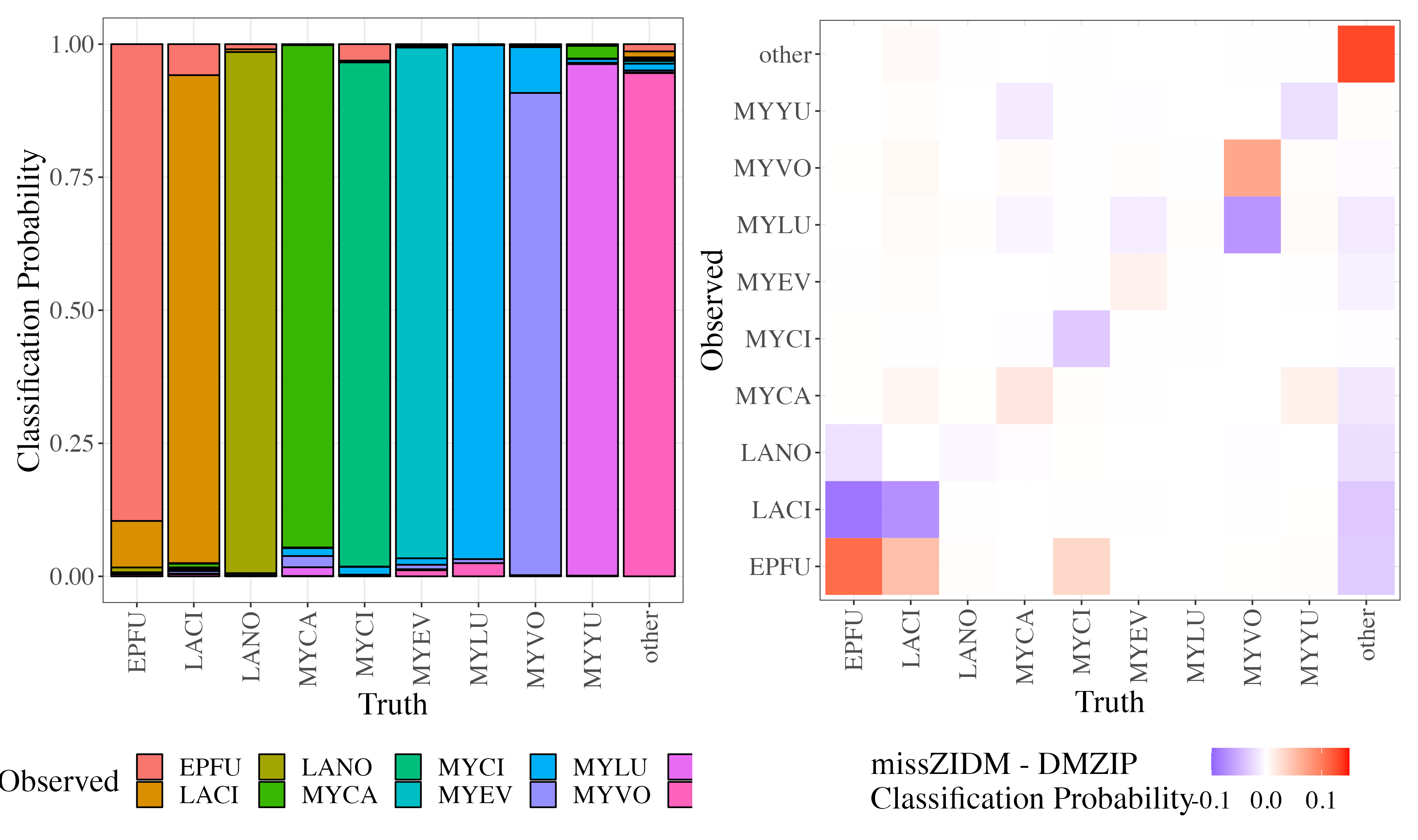}
\caption{ \textcolor{black}{Bat Monitoring Application Results:} The left subplot shows the estimated confusion matrix probabilities using the proposed  missZIDM model. The right subplot presents a heatmap of the difference between the estimated confusion matrix probabilities of missZIDM and DMZIP.}   
\label{fig:classification:models}
\end{figure}

  Figure \ref{fig:classification:models} presents the estimated confusion matrix probabilities for the proposed missZIDM model, as well as the differences with the DMZIP model. The models found relatively similar results, with the average (SD) absolute difference between all cells of the confusion matrix equal to 1\% (2\%). The largest differences in misclassification ($1 - $diagonal elements of the confusion matrix) estimates were 
 DMZIP estimating 10.1\% more misclassification for  EPFU, 7\% more for MYVO, and 13\% more for the \textit{other} category. Additionally, the proposed method estimated 7\% more misclassification for LACI and 3\% more for MYYU compared to DMZIP. In the Supplementary Material, we provide tables for the estimated confusion matrix probabilities and corresponding 95\% credible intervals for missZIDM and DMZIP (Supplementary Tables S7 and S8, respectively).

\begin{figure}[ht]
\includegraphics[width=\textwidth]{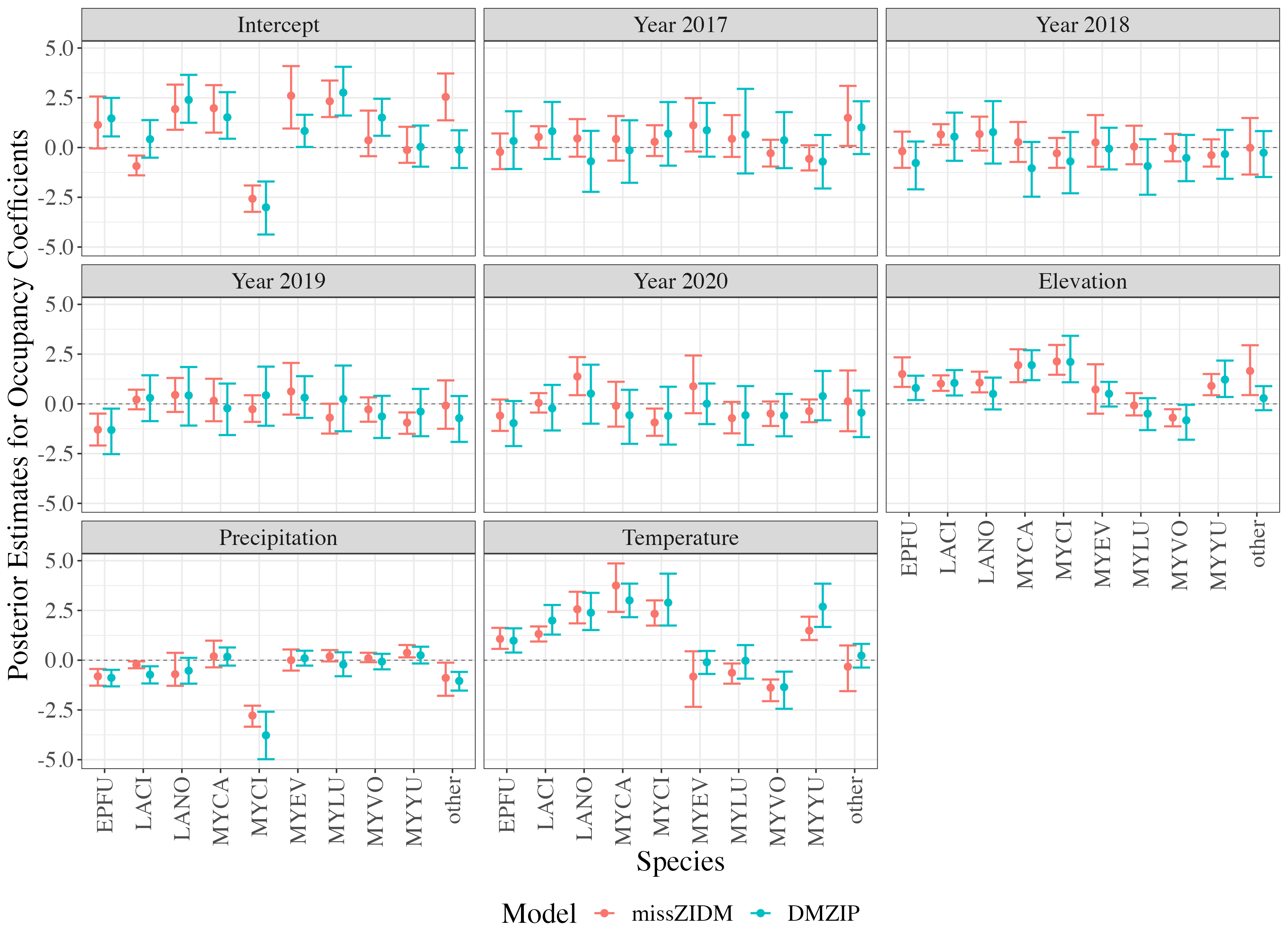}
\caption{\textcolor{black}{Bat Monitoring Application Results:} Posterior estimates of regression coefficients for occupancy for the proposed missZIDM and DMZIP models. Dot represents the posterior mean with error bars representing the 95\% credible intervals. }
\label{fig:occupancy}
\end{figure}

 We investigated the estimated covariate associations for occupancy in both models. The estimates plotted in Figure \ref{fig:occupancy} for  $\boldsymbol{\beta}_{\eta}$ 
are interpreted as log odds ratios for occupancy at each site visit. Overall, the results were quite similar between the models. Neither method found a strong effect for time on occupancy. However, both models estimated a decrease in the odds of occupancy for EPFU in 2019 compared to 2016, and DMZIP estimated an increase in the odds of occupancy for LANO in 2020 compared to 2016. We found that an increase in elevation was associated with an increase in the odds of occupancy for most species, with the exception of MYVO. We observed mostly negative associations between precipitation and occupancy, although most of the 95\% credible intervals contained 0 for both models. The strongest relation was for MYCI, where a millimeter increase in precipitation was associated with a 95\% decrease in occupancy. EPFU, LACI, LANO, MYCA, MYCI, and MYYU were all found to have positive associations between temperature and occupancy. However, temperature was negatively associated with occupancy for MYVO.  Temperature has previously been found to be associated with occupancy for EPFU, LACI, and LANO in a study conducted in North Carolina during the winter season \citep{parker2020species}. Typically the proposed method was more conservative than DMZIP with respect to parameter uncertainty for all covariate effects.  

\begin{figure}[ht]
\includegraphics[width=\textwidth]{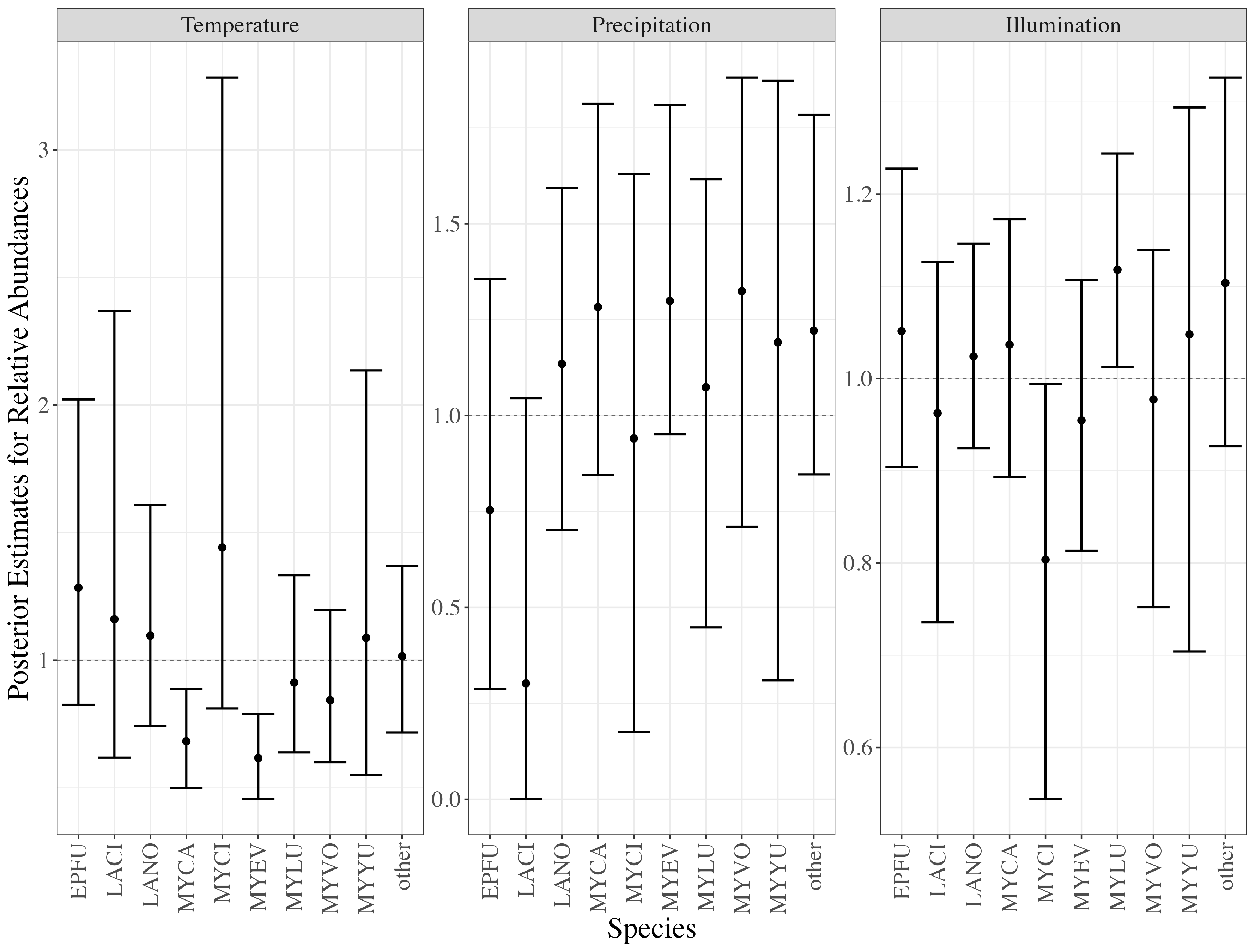}
\caption{\textcolor{black}{Bat Monitoring Application Results:} Posterior estimates of the multiplicative effect for a one-unit increase in each covariate on the relative abundance of each species using the proposed model. Dot represents the posterior mean with error bars representing the 95\% credible intervals. }
\label{fig:RA}
\end{figure}

\begin{figure}[H]
\includegraphics[width=\textwidth]{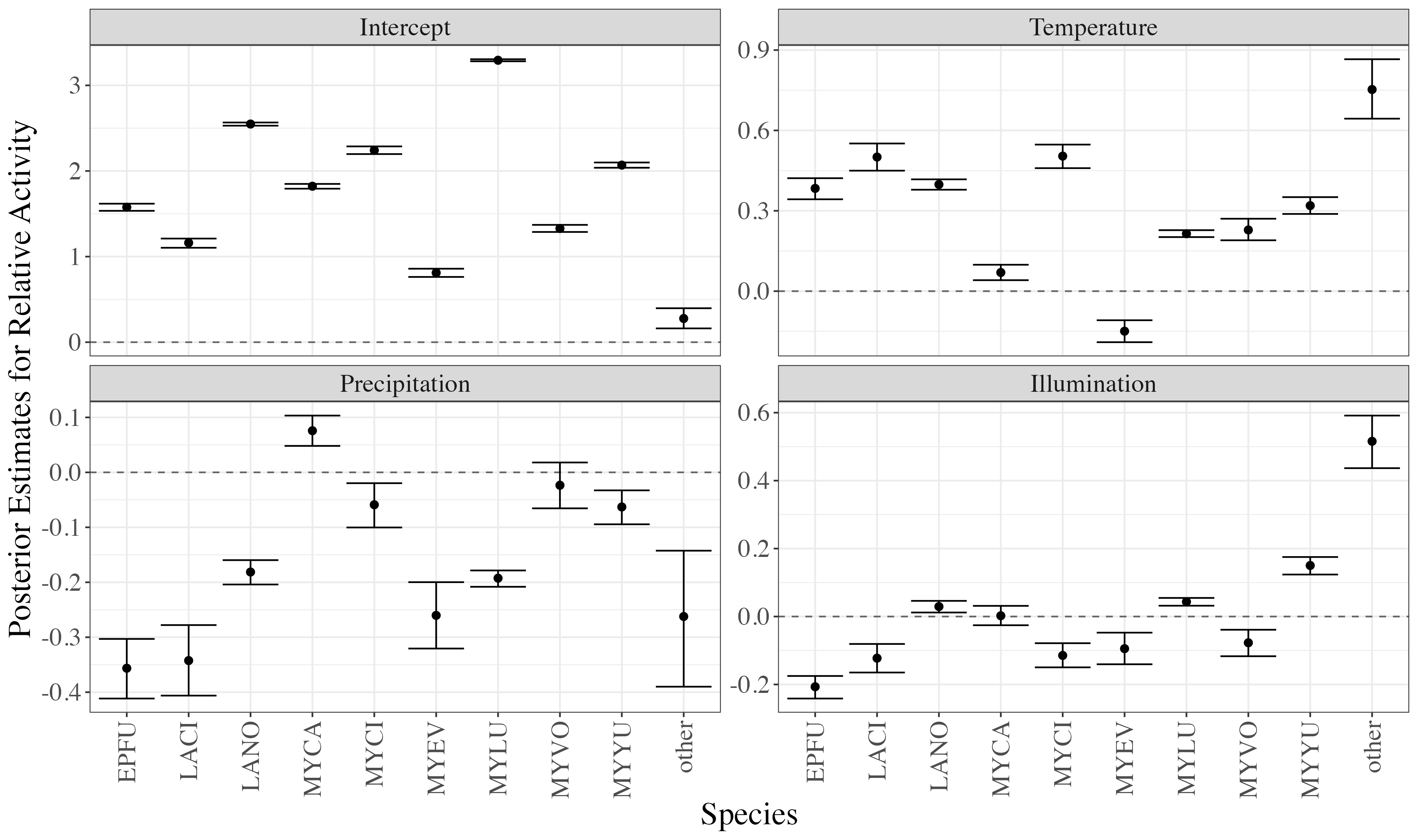}
\caption{\textcolor{black}{Bat Monitoring Application Results:} Posterior estimates of regression coefficients for relative activity using the DMZIP model. Dot represents the posterior mean with error bars representing the 95\% credible intervals. }
\label{fig:activity}
\end{figure}

 Figure \ref{fig:RA} presents the estimated multiplicative effect for a one-unit increase in each covariate (i.e., temperature, precipitation, and illumination) on the relative abundance of each species using the proposed model given the sample average of the other covariates, $\pi_{tp}$. As described previously, the effect of a covariate on each species' relative abundance depends on its association with the other species' relative abundances. As such, a positive (negative) association between a covariate and a species' relative \textit{activity} does not imply a positive (negative) association with the species' relative \textit{abundance}. For example, precipitation was found to be negatively associated with a majority of the bat species' relative activity using the DMZIP model (Figure \ref{fig:activity}) as well as the concentration parameters of the proposed model (Supplementary Figure S8). However, precipitation was positively associated with 7/10 species' relative abundances, which reflects the differences in the effects of precipitation on bat activity across species. The overall negative relations observed between bat activity and precipitation in this analysis corroborate previous findings that attribute the reduction to various factors, including increased flight metabolism of wet bats, interference with echolocation, and availability of prey  \citep{griffin1971importance,burles2009influence,voigt2011rain}. Previous studies have shown that activity by insectivorous bats is sensitive to environmental conditions, with different effects observed for different species \citep{thies2006influence,vasquez2020species,rodriguez2024species}. For example, bat species' response to moonlight intensity and temperature has been found to be species-specific \citep{saldana2013lunar,klug2016environmental,appel2017aerial,vasquez2020species}. We found mixed results regarding the associations between nightly minimum air temperature and moon illumination and relative abundances. We observed a negative association between temperature and the relative abundances of  MYCA and MYEV. We also observed a negative association between moon illumination and the relative abundance of MYCI with missZIDM, in addition to its relative activity with DMZIP.  

 One of the advantages of the proposed method is that it does not require making distributional assumptions for the latent ecological process to simultaneously model non-detection and misclassification in multispecies settings. By leveraging a combination of data augmentation techniques, the proposed method is able to scale to high-dimensional settings while additionally modeling the latent classifications and generating inference on the true relative abundances of each species at each site measurement. While missZIDM teases apart the task of modeling the true classifications and ecological process, it does not preclude embedding the proposed model into larger hierarchical frameworks aimed at simultaneously inferring the ecological process. This is similar to techniques used in integrated population modeling \citep{schaub2011integrated}.



\clearpage

\subsection{ \textcolor{black}{Application to Official Crime Data}}\label{sec::crime}

\textcolor{black}{In this analysis, we apply the proposed method to incident-level official crime data  that include victim-involved offenses reported by Colorado law enforcement agencies to the National Incident-Based Reporting System (NIBRS) in 2022 \textcolor{black}{(publicly available online; \cite{ICPSR})}. Collected under the Uniform Crime Reporting Program (UCR), NIBRS data   contain criminal incidents that are known to law enforcement and serve as the national standard for official crime reporting in the United States \citep{addington2019nibrs}.}
\textcolor{black}{Since the UCR inception in 1991, NIBRS data have continued to improve upon official criminal data monitoring by providing  greater breadth and depth in reporting information about criminal incidents, context about specific crime problems, and  more capacity for data analysis about nation-wide crime trends.}  

 \textcolor{black}{One of the challenges of analyzing NIBRS crime data is \textcolor{black}{the high occurrence of zero counts} due to rarely occurring crimes (e.g., murder) \citep{rydberg2017utilizing, luo2022associating}, 
data entry errors \citep{wheeler2018monitoring}, and the underreporting of criminal incidents \citep{skogan1974validity,wormeli2018criminal}. For example, victim-involved criminal incidents including violent, domestic violence, or sexual assault offenses are particularly likely to go unreported to law enforcement \citep{langton2017second}. Another challenge of modeling NIBRS crime data is they are subject to misclassification  by law enforcement agencies \citep{bibel2015considerations}. Common misclassification errors may occur when a criminal offense is coded incorrectly in comparison to qualitative incident information provided in the probable cause affidavit \citep{nolan2011estimating}. For example, a criminal incident may be \textcolor{black}{incorrectly} coded to include a simple assault offense \textcolor{black}{instead of an aggravated assault} when the presence of a weapon is detailed in the affidavit. Misclassification errors may also occur when the NIBRS coding category fails to include the necessary level of specificity for a certain type of offense circumstance within a criminal incident \citep{osborne2019utilizing, haberman2022robbery}. For example, incidents that involve a robbery offense cannot be differentiated further by typology such as carjacking, bank robbery, or residential invasion. 
The amount of misclassification has been found to vary by crime type \citep{nolan2011estimating} and can result in the over- and undercounting of certain crime categories. Without accurate reporting, criminal justice research using official data \textcolor{black}{is} vulnerable to \textcolor{black}{ potential  biases introduced by law enforcement reporting practices that misrepresent true criminal phenomenon} \citep{pina2023impact}. Given the widespread use of NIBRS data to inform resource allocation, criminal justice policy, and the empirical understanding of crime phenomena, it is critical to account for potential classification errors to improve the accuracy of criminal incident reporting. }


\textcolor{black}{In this analysis, we demonstrate how the proposed method can be used to account for potential zero-inflation and misclassification when estimating crime incidents. We applied the proposed method to incident-level victim-involved crime data (i.e., murder, rape, sodomy, sexual assault with an object (SAO), fondling, robbery, aggravated assault, simple assault) reported by all Colorado agencies to NIBRS in 2022. These data capture 55,198 total reported incidents over $N=216$ agencies, ranging from 318 incidences of murder to 29,747 incidents of simple assault. \textcolor{black}{We observed 50\% of the crime categories reported by law enforcement agencies were zero counts. Across the different categories, murder had the highest proportion of zero counts (i.e., 76\%).}}

 \textcolor{black}{In the absence of validation data to inform misclassification probabilities, we evaluated the model using three different prior specifications. In the first setting (\textit{Naive}), we assume that the probability of an at-risk observation for a given crime, $\eta_{t}$, is 0.50, there is a 0.05 probability of misclassification for each crime (i.e., $\psi_{t} = 0.05$), and each crime is equally likely to be misclassified as another crime with a small probability of a lucky guess (i.e., $\nu_{tc} = 1$ with $\nu_{tt} = 0.001$). Without knowledge regarding the amount of misclassification in the data, it is difficult to specify the concentration hyperparameters for the true relative abundances. Thus in the \textit{Naive} setting, we take an empirical Bayes approach and set the concentration hyperparameters to the log of the average observed relative abundances across agencies scaled by 1,000.}
 
 \textcolor{black}{The assumption that each crime is equally likely to be misclassified as another crime is somewhat unrealistic as some crimes are more similar than others. For example, a simple assault is more likely to be misclassified as an aggravated assault than a homicide. Therefore in a second setting (\textit{Blocked}), we formed three groups of victim-involved crimes including murder, sex offenses (i.e., rape, sodomy, SAO, and fondling), and other violent crimes (i.e., robbery, aggravated assault, and simple assault) within which misclassification was more likely. For implementation, we  assumed a block matrix for $\boldsymbol{\nu}$ with $\nu_{tc} = $ 1,000 for the off-diagonal elements within each block and $\nu_{tc} = 1\mathrm{e}{-5}$, otherwise. The other hyperparameters were set similar to  the \textit{Naive} model.}
 
 \textcolor{black}{In a third setting, we used historical data from \cite{nolan2011estimating} to inform the model (\textit{Historical}).   Previously, \cite{nolan2011estimating} performed a validation study on 3 of the 12 largest municipal police agencies in a mostly rural southeastern state for 15 crime categories using Uniform Crime Reporting (UCR) data collected in 2002. As this information may not fully represent those found in our application data, we recommend validating the results obtained from this analysis prior to generalization.     In the validation data, \cite{nolan2011estimating} found that  21.6\% of the rape incidents,  7.3\% of the other sex offenses, 5.0\% of the robberies,  8.4\% of the aggravated assaults, and 1.8\% of the simple assaults were misclassified. We used these misclassification probabilities to specify $\psi_{t}$ in our model, with 7.3\% misclassification assumed for each of the sex offense incidents analyzed.  Because there were no misclassifications found in the validation set for murder, we \textcolor{black}{assumed} a small probability of error for analysis (i.e., 0.0001). Additionally, we set the concentration hyperparameters for the true classifications to the log of the ``statistically adjusted'' relative abundances from \cite{nolan2011estimating} scaled by 1,000. We assumed a similar block  matrix for $\boldsymbol{\nu}$ with $\nu_{tc} =$ 1,000 for the off-diagonal elements within each block, $\nu_{tt} = 0.001$, and $\nu_{tc} = 1\mathrm{e}{-8}$, otherwise. The remaining priors were specified similar to the other settings.}

\textcolor{black}{In each setting, the MCMC algorithm was run for 200,000 iterations, thinning to every 25th iteration and treating the first 2,000 as burn-in, leaving 2,000 iterations for inference. Estimated true classifications were obtained using the \texttt{salso} method with Binder's loss \citep{dahl2022search}. Convergence and mixing of the models was visually inspected using traceplots. See Supplementary Material for traceplots for a random subset of the parameters (Supplementary Figures S9-S12).}

\begin{figure}[H]
\centerline{\includegraphics[width=1.35\textwidth]{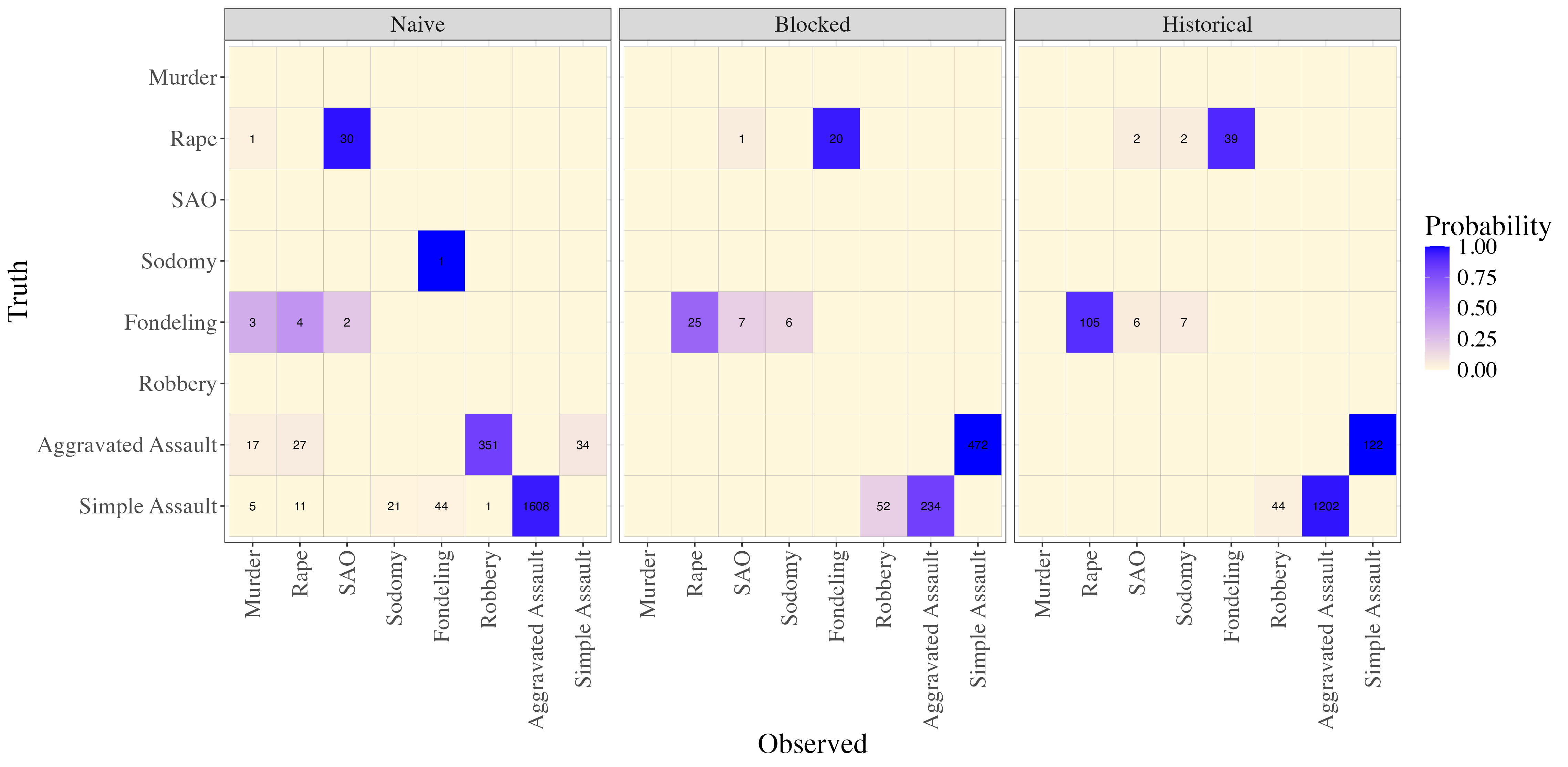}}
\caption{\textcolor{black}{Crime Application Results: Probability of observed classification given the true classification estimated using the three modeling assumptions. Note that each row sums to one, and the counts represent the number of misclassified incidents.}}
\label{fig:CrimeClass}
\end{figure}

\textcolor{black}{Table \ref{tab::counts} reports the estimated mean incidence with 95\% credible intervals for each crime using the proposed method. Under the assumptions of the \textit{Naive} model, the misclassified incidents from the smaller crime categories gravitated towards the largest category, simple assault.  \textcolor{black}{Overall, we observed the estimated crime incidents for the \textit{Blocked} model were the most similar to the observed classifications. These results were expected as the prior specification in this setting aligns closely to the observed data.} The \textit{Naive} model  estimated the highest number of misclassifications (2,160), compared to 817 and 1,529 for the \textit{Blocked} and \textit{Historical} models, respectively (Figure \ref{fig:CrimeClass}). With the \textit{Blocked} and \textit{Historical} models, no misclassifications were observed outside of the blocked structure, whereas the \textit{Naive} model estimated 134 misclassifications. In all settings, the majority of misclassifications estimated were for true simple assaults  observed as aggravated assaults.} 
 
 \textcolor{black}{In this analysis, we demonstrate how the proposed method can be used to account for potential biases in the estimation of criminal incidents attributable to zero-inflation in and misclassification of official crime data. While we illustrated how prior knowledge and historical data can be used to inform the model, using non-representative or inaccurate information may do more harm than good mitigating biases in official crime data. \textcolor{black}{However, specifying informative priors  based on historical data or expert knowledge may still be preferred over naive prior assumptions when appropriate. For example in our application, it is reasonable to justify that murders are rarely misclassified. Whereas the true crime classification probabilities obtained from the historical data collected in a mostly rural southeastern state in 2002 may not be fully representative of the crime patterns found in Colorado in 2022.} In practice, we recommend collecting validation data for a subset of the data analyzed to account for spatial and temporal variation in misclassfication rates.  }


\begin{sidewaystable}[] 
\textcolor{black}{
\footnotesize \setlength{\tabcolsep}{1pt} 
\label{tab::counts}
\begin{tabular}{ccccccccc}
\hline
                            & Murder     & Rape         & SAO        & Sodomy     & Fondling     & Robbery      & Aggravated Assault & Simple Assault \\ \hline
Observed                    & 318        & 2245         & 620        & 579        & 2340         & 4215         & 15134              & 29747          \\
\multirow{2}{*}{Naive}      & 260.2      & 2156.7       & 546.1      & 579.8      & 2273.8       & 3998.7       & 14195.0            & 31187.7        \\
                            & (247, 277) & (2125, 2194) & (522, 580) & (566, 598) & (2193, 2334) & (3950, 4050) & (14119, 14269)     & (31107, 31275) \\
\multirow{2}{*}{Blocked}    & 313.6      & 2236.1       & 622.6      & 584.6      & 2340.7       & 4268.0       & 15385.8            & 29446.6        \\
                            & (257, 318) & (2230, 2242) & (617, 628) & (579, 591) & (2334, 2346) & (4249, 4288) & (15337, 15448)     & (29381, 29493) \\
\multirow{2}{*}{Historical} & 344.8      & 2169.7       & 644.9      & 590.0      & 2352.6       & 4290.9       & 14073.7            & 30731.5        \\
                            & (336, 354) & (2160, 2178) & (635, 656) & (582, 599) & (2341, 2363) & (4271, 4312) & (13975, 14184)     & (30623, 30832) \\ \hline
\end{tabular} 
\caption{Crime Application Results: Observed and estimated mean incidents and corresponding 95\% credible intervals \textcolor{black}{(below in parentheses)} obtained with the proposed model using different prior formulations.  } }
\end{sidewaystable}

\section{Conclusions}

In this work, we propose the first method for simultaneously accommodating false positives and false negatives in multinomial data. Our model can naturally incorporate existing knowledge of zero-inflation or misclassification through prior specification and/or easily accommodate validation data to inform the model. In simulation, we demonstrated that our approach obtains similar or improved estimation performance for at-risk, true classification, and misclassification probabilities compared to alternative methods that ignore one or both forms of measurement error.  We further show how the proposed approach can provide more accurate estimation for at-risk and classification probabilities than existing multivariate methods in the presence of overdispersed data because it does not require accurately specifying the latent count process. By simultaneously modeling misclassification at the individual level with potential non-detection, our approach accommodates the uncertainty of non-detection occurring if the species is present but not observed at each site visit and if each of the individuals observed for a particular species at a given site are incorrectly classified. Conceptually, the method of \cite{wright2020modelling} and \cite{spiers2022estimating} could be adjusted to accommodate potential overdispersion by replacing the Poisson distribution with a negative binomial distribution.  This approach serves as a potentially viable alternative for modeling misclassification in overdispersed multivariate count data, but it would not be appropriate for multinomial or compositional data settings where the total number of counts is fixed.

  \textcolor{black}{While demonstrated in ecological and criminal justice research settings}, the proposed method is applicable to other settings in which zero-inflated multinomial data with potential misclassification are collected. For example, the method could be used to model zero-inflated multivariate count data collected in human microbiome research settings that are subject to measurement error introduced at various stages of the measurement protocol \citep{pollock2018madness,clausen2022evaluating}. Additionally, citizen science and crowdsourcing projects often task contributors with classifying different features, such as radio technosignatures to help detect extraterrestrial life  \citep{margot2019radio}, types of stars based on their spectra \citep{SCOPE}, as well as images, videos, sounds, water samples, and/or sensor data for biodiversity research, Earth observation, and geography and climate change research \citep{pocock2014choosing, fraisl2022citizen, ficetola2016limit, schmidt2013site, willoughby2016importance, lahoz2016statistical}. In the context of conservation research and wildlife monitoring studies, the proposed method could be customized to answer pressing research questions. For example, to accommodate occupancy dynamics, one could assume the probability a site is in a given occupancy state is governed by a Markov process, similar to \cite{miller2013determining}. \textcolor{black}{In the second application study, we investigated official data of criminal  incidents  reported by law enforcement agencies. However, it is well known that there are discrepancies between official, victim- and self-reported crime data. In future work, our proposed method could be used to evaluate potential biases in official data reported across different sources by accommodating multiple measurements for each site, or in this case, jurisdiction, and covariate information for true relative abundances.}

\begin{supplement}
\stitle{Supplementary Material}
\sdescription{The Supplementary Material contains detailed derivations of the MCMC algorithm, sensitivity analysis, and additional tables and figures. 
}
\end{supplement}
\begin{supplement}
 \stitle{missZIDM \texttt{R} package}
\sdescription{  This file contains code to generate data similar to the simulation study as well as a vignette demonstrating how to apply the method and perform inference.}  
\end{supplement}

\clearpage
\begin{acks}[Acknowledgments]
MDK gratefully acknowledges the support of NSF grant DMS-2245492. AK gratefully acknowledges the support of NSF grants DMS-2330089 and SES-2338428. The opinions, findings, and conclusions
expressed are those of the authors and do not
necessarily reflect the views of the NSF. The authors thank Dr. Kathi Irvine for providing useful discussions about modeling bat acoustic data.
\end{acks}

  \bibliographystyle{apalike} 
 \bibliography{bib.bib}

\begin{thebibliography}{}

\bibitem[Addington, 2019]{addington2019nibrs}
Addington, L.~A. (2019).
\newblock {NIBRS} as the new normal: {W}hat fully incident-based crime data
  mean for researchers.
\newblock {\em Handbook on Crime and Deviance}, pages 21--33.

\bibitem[Aitchison and Ho, 1989]{aitchison1989multivariate}
Aitchison, J. and Ho, C. (1989).
\newblock The multivariate {P}oisson-log normal distribution.
\newblock {\em Biometrika}, 76(4):643--653.

\bibitem[Appel et~al., 2017]{appel2017aerial}
Appel, G., L{\'o}pez-Baucells, A., Ernest-Magnusson, W., and Bobrowiec, P.
  E.~D. (2017).
\newblock Aerial insectivorous bat activity in relation to moonlight intensity.
\newblock {\em Mammalian Biology}, 85:37--46.

\bibitem[Bayley and Peterson, 2001]{bayley2001approach}
Bayley, P.~B. and Peterson, J.~T. (2001).
\newblock An approach to estimate probability of presence and richness of fish
  species.
\newblock {\em Transactions of the American Fisheries Society},
  130(4):620--633.

\bibitem[Bibel, 2015]{bibel2015considerations}
Bibel, D. (2015).
\newblock Considerations and cautions regarding {NIBRS} data: {A} view from the
  field.
\newblock {\em Justice Research and Policy}, 16(2):185--194.

\bibitem[Blasco-Moreno et~al., 2019]{blasco2019does}
Blasco-Moreno, A., P{\'e}rez-Casany, M., Puig, P., Morante, M., and Castells,
  E. (2019).
\newblock What does a zero mean? {U}nderstanding false, random and structural
  zeros in ecology.
\newblock {\em Methods in Ecology and Evolution}, 10(7):949--959.

\bibitem[Broms et~al., 2015]{broms2015accounting}
Broms, K.~M., Hooten, M.~B., and Fitzpatrick, R.~M. (2015).
\newblock Accounting for imperfect detection in {H}ill numbers for biodiversity
  studies.
\newblock {\em Methods in Ecology and Evolution}, 6(1):99--108.

\bibitem[Brooks and Gelman, 1998]{brooks1998general}
Brooks, S.~P. and Gelman, A. (1998).
\newblock General methods for monitoring convergence of iterative simulations.
\newblock {\em Journal of Computational and Graphical Statistics},
  7(4):434--455.

\bibitem[{Bureau of Justice Statistics}, 2023]{ICPSR}
{Bureau of Justice Statistics} (2023).
\newblock {National Incident-Based Reporting System, 2022: Extract Files.
  Inter-university Consortium for Political and Social Research}.
\newblock \url{https://doi.org/10.3886/ICPSR38925.v1}.
\newblock Accessed: 2024-06-22.

\bibitem[Burles et~al., 2009]{burles2009influence}
Burles, D., Brigham, R., Ring, R., and Reimchen, T. (2009).
\newblock Influence of weather on two insectivorous bats in a temperate
  {P}acific {N}orthwest rainforest.
\newblock {\em Canadian Journal of Zoology}, 87(2):132--138.

\bibitem[Chambert et~al., 2018a]{chambert2018two}
Chambert, T., Grant, E. H.~C., Miller, D.~A., Nichols, J.~D., Mulder, K.~P.,
  and Brand, A.~B. (2018a).
\newblock Two-species occupancy modelling accounting for species
  misidentification and non-detection.
\newblock {\em Methods in Ecology and Evolution}, 9(6):1468--1477.

\bibitem[Chambert et~al., 2015]{chambert2015modeling}
Chambert, T., Miller, D.~A., and Nichols, J.~D. (2015).
\newblock Modeling false positive detections in species occurrence data under
  different study designs.
\newblock {\em Ecology}, 96(2):332--339.

\bibitem[Chambert et~al., 2018b]{chambert2018new}
Chambert, T., Waddle, J.~H., Miller, D.~A., Walls, S.~C., and Nichols, J.~D.
  (2018b).
\newblock A new framework for analysing automated acoustic species detection
  data: {O}ccupancy estimation and optimization of recordings post-processing.
\newblock {\em Methods in Ecology and Evolution}, 9(3):560--570.

\bibitem[Chiquet et~al., 2021]{chiquet2021poisson}
Chiquet, J., Mariadassou, M., and Robin, S. (2021).
\newblock The {P}oisson-lognormal model as a versatile framework for the joint
  analysis of species abundances.
\newblock {\em Frontiers in Ecology and Evolution}, 9:188.

\bibitem[Clausen and Willis, 2022]{clausen2022evaluating}
Clausen, D.~S. and Willis, A.~D. (2022).
\newblock Evaluating replicability in microbiome data.
\newblock {\em Biostatistics}, 23(4):1099--1114.

\bibitem[Copas and Hilton, 1990]{copas1990record}
Copas, J. and Hilton, F. (1990).
\newblock Record linkage: {S}tatistical models for matching computer records.
\newblock {\em Journal of the Royal Statistical Society: Series A (Statistics
  in Society)}, 153(3):287--312.

\bibitem[Dahl et~al., 2022]{dahl2022search}
Dahl, D.~B., Johnson, D.~J., and M{\"u}ller, P. (2022).
\newblock Search algorithms and loss functions for {B}ayesian clustering.
\newblock {\em Journal of Computational and Graphical Statistics},
  31(4):1189--1201.

\bibitem[Dai et~al., 2019]{dai2019batch}
Dai, Z., Wong, S.~H., Yu, J., and Wei, Y. (2019).
\newblock Batch effects correction for microbiome data with
  {D}irichlet-multinomial regression.
\newblock {\em Bioinformatics}, 35(5):807--814.

\bibitem[Datta et~al., 2021]{datta2021regularized}
Datta, A., Fiksel, J., Amouzou, A., and Zeger, S.~L. (2021).
\newblock Regularized {B}ayesian transfer learning for population-level
  etiological distributions.
\newblock {\em Biostatistics}, 22(4):836--857.

\bibitem[DeLisle and Barker, 2024]{SCOPE}
DeLisle, T. and Barker, T. (2024).
\newblock {SCOPE} stellar classification online public exploration.
\newblock \url{http://scope.pari.edu/}.
\newblock Accessed: 2024-03-24.

\bibitem[Devarajan et~al., 2020]{devarajan2020multi}
Devarajan, K., Morelli, T.~L., and Tenan, S. (2020).
\newblock Multi-species occupancy models: Review, roadmap, and recommendations.
\newblock {\em Ecography}, 43(11):1612--1624.

\bibitem[Dorazio et~al., 2011]{dorazio2011modern}
Dorazio, R.~M., Gotelli, N.~J., and Ellison, A.~M. (2011).
\newblock Modern methods of estimating biodiversity from presence-absence
  surveys.
\newblock {\em Biodiversity Loss in a Changing Planet}, pages 277--302.

\bibitem[Dorazio et~al., 2006]{dorazio2006estimating}
Dorazio, R.~M., Royle, J.~A., S{\"o}derstr{\"o}m, B., and Glimsk{\"a}r, A.
  (2006).
\newblock Estimating species richness and accumulation by modeling species
  occurrence and detectability.
\newblock {\em Ecology}, 87(4):842--854.

\bibitem[Eddelbuettel and Fran{\c{c}}ois, 2011]{eddelbuettel2011rcpp}
Eddelbuettel, D. and Fran{\c{c}}ois, R. (2011).
\newblock Rcpp: Seamless {R} and {C}++ integration.
\newblock {\em Journal of Statistical Software}, 40:1--18.

\bibitem[Ficetola et~al., 2016]{ficetola2016limit}
Ficetola, G., Taberlet, P., and Coissac, E. (2016).
\newblock How to limit false positives in environmental {DNA} and
  metabarcoding?
\newblock {\em Molecular Ecology Resources}, 16(3):604--607.

\bibitem[Fraisl et~al., 2022]{fraisl2022citizen}
Fraisl, D., Hager, G., Bedessem, B., Gold, M., Hsing, P.-Y., Danielsen, F.,
  Hitchcock, C.~B., Hulbert, J.~M., Piera, J., Spiers, H., et~al. (2022).
\newblock Citizen science in environmental and ecological sciences.
\newblock {\em Nature Reviews Methods Primers}, 2(1):64.

\bibitem[Fr{\'e}nay and Verleysen, 2013]{frenay2013classification}
Fr{\'e}nay, B. and Verleysen, M. (2013).
\newblock Classification in the presence of label noise: {A} survey.
\newblock {\em IEEE Transactions on Neural Networks and Learning Systems},
  25(5):845--869.

\bibitem[Griffin, 1971]{griffin1971importance}
Griffin, D.~R. (1971).
\newblock The importance of atmospheric attenuation for the echolocation of
  bats (chiroptera).
\newblock {\em Animal Behaviour}, 19(1):55--61.

\bibitem[Guillera-Arroita et~al., 2017]{guillera2017dealing}
Guillera-Arroita, G., Lahoz-Monfort, J.~J., van Rooyen, A.~R., Weeks, A.~R.,
  and Tingley, R. (2017).
\newblock Dealing with false-positive and false-negative errors about species
  occurrence at multiple levels.
\newblock {\em Methods in Ecology and Evolution}, 8(9):1081--1091.

\bibitem[Haberman et~al., 2022]{haberman2022robbery}
Haberman, C.~P., Clutter, J.~E., and Lee, H. (2022).
\newblock A robbery is a robbery is a robbery? {E}xploring crime specificity in
  official police incident data.
\newblock {\em Police Practice and Research}, 23(4):429--443.

\bibitem[Hoeting et~al., 2000]{hoeting2000improved}
Hoeting, J.~A., Leecaster, M., and Bowden, D. (2000).
\newblock An improved model for spatially correlated binary responses.
\newblock {\em Journal of Agricultural, Biological, and Environmental
  Statistics}, pages 102--114.

\bibitem[Jasra et~al., 2005]{jasra2005markov}
Jasra, A., Holmes, C., and Stephens, D. (2005).
\newblock Markov chain {M}onte {C}arlo {M}ethods and the label switching
  problem in {B}ayesian mixture modeling.
\newblock {\em Statistical Science}, 20(1):50--67.

\bibitem[Jiang et~al., 2021]{jiang2021bayesian}
Jiang, S., Xiao, G., Koh, A.~Y., Kim, J., Li, Q., and Zhan, X. (2021).
\newblock A {B}ayesian zero-inflated negative binomial regression model for the
  integrative analysis of microbiome data.
\newblock {\em Biostatistics}, 22(3):522--540.

\bibitem[Kellner and Swihart, 2014]{kellner2014accounting}
Kellner, K.~F. and Swihart, R.~K. (2014).
\newblock Accounting for imperfect detection in ecology: {A} quantitative
  review.
\newblock {\em PloS One}, 9(10):e111436.

\bibitem[Kl{\"u}g-Baerwald et~al., 2016]{klug2016environmental}
Kl{\"u}g-Baerwald, B.~J., Gower, L.~E., Lausen, C., and Brigham, R. (2016).
\newblock Environmental correlates and energetics of winter flight by bats in
  southern {A}lberta, {C}anada.
\newblock {\em Canadian Journal of Zoology}, 94(12):829--836.

\bibitem[Koslovsky, 2023]{koslovsky2023bayesian}
Koslovsky, M.~D. (2023).
\newblock A {B}ayesian zero-inflated {D}irichlet-multinomial regression model
  for multivariate compositional count data.
\newblock {\em Biometrics}.

\bibitem[Koslovsky et~al., 2020]{koslovsky2020microbiome}
Koslovsky, M.~D., Hoffman, K.~L., Daniel, C.~R., and Vannucci, M. (2020).
\newblock A {B}ayesian model of microbiome data for simultaneous identification
  of covariate associations and prediction of phenotypic outcomes.
\newblock {\em The Annals of Applied Statistics}, 14(3):1471--1492.

\bibitem[Lahoz-Monfort et~al., 2016]{lahoz2016statistical}
Lahoz-Monfort, J.~J., Guillera-Arroita, G., and Tingley, R. (2016).
\newblock Statistical approaches to account for false-positive errors in
  environmental {DNA} samples.
\newblock {\em Molecular Ecology Resources}, 16(3):673--685.

\bibitem[Langton et~al., 2017]{langton2017second}
Langton, L., Planty, M., and Lynch, J.~P. (2017).
\newblock Second major redesign of the {N}ational {C}rime {V}ictimization
  {S}urvey {(NCVS)}.
\newblock {\em Criminology \& Public Policy}, 16:1049.

\bibitem[Lele et~al., 2012]{lele2012dealing}
Lele, S.~R., Moreno, M., and Bayne, E. (2012).
\newblock Dealing with detection error in site occupancy surveys: {W}hat can we
  do with a single survey?
\newblock {\em Journal of Plant Ecology}, 5(1):22--31.

\bibitem[Loeb et~al., 2015]{loeb2015plan}
Loeb, S., Rodhouse, T., Ellison, L., Lausen, C., Reichard, J., Irvine, K.,
  Ingersoll, T., Coleman, J., Thogmartin, W., Sauer, J., et~al. (2015).
\newblock A plan for the {N}orth {A}merican {B}at {M}onitoring {P}rogram
  {(NABat)}.
\newblock {\em General Technical Report-Southern Research Station, USDA Forest
  Service}.

\bibitem[Luo et~al., 2022]{luo2022associating}
Luo, L., Deng, M., Shi, Y., Gao, S., and Liu, B. (2022).
\newblock Associating street crime incidences with geographical environment in
  space using a zero-inflated negative binomial regression model.
\newblock {\em Cities}, 129:103834.

\bibitem[MacKenzie et~al., 2003]{mackenzie2003estimating}
MacKenzie, D.~I., Nichols, J.~D., Hines, J.~E., Knutson, M.~G., and Franklin,
  A.~B. (2003).
\newblock Estimating site occupancy, colonization, and local extinction when a
  species is detected imperfectly.
\newblock {\em Ecology}, 84(8):2200--2207.

\bibitem[MacKenzie et~al., 2002]{mackenzie2002estimating}
MacKenzie, D.~I., Nichols, J.~D., Lachman, G.~B., Droege, S., Andrew~Royle, J.,
  and Langtimm, C.~A. (2002).
\newblock Estimating site occupancy rates when detection probabilities are less
  than one.
\newblock {\em Ecology}, 83(8):2248--2255.

\bibitem[MacKenzie et~al., 2017]{mackenzie2017occupancy}
MacKenzie, D.~I., Nichols, J.~D., Royle, J.~A., Pollock, K.~H., Bailey, L., and
  Hines, J.~E. (2017).
\newblock {\em Occupancy estimation and modeling: inferring patterns and
  dynamics of species occurrence}.
\newblock Elsevier.

\bibitem[Margot et~al., 2019]{margot2019radio}
Margot, J.-L., Croft, S., Lazio, J., Tarter, J., and Korpela, E. (2019).
\newblock The radio search for technosignatures in the decade 2020—2030.
\newblock {\em Bulletin of the American Astronomical Society}, 51(3):298.

\bibitem[McClintock et~al., 2010]{mcclintock2010unmodeled}
McClintock, B.~T., Bailey, L.~L., Pollock, K.~H., and Simons, T.~R. (2010).
\newblock Unmodeled observation error induces bias when inferring patterns and
  dynamics of species occurrence via aural detections.
\newblock {\em Ecology}, 91(8):2446--2454.

\bibitem[Miller et~al., 2013]{miller2013determining}
Miller, D.~A., Nichols, J.~D., Gude, J.~A., Rich, L.~N., Podruzny, K.~M.,
  Hines, J.~E., and Mitchell, M.~S. (2013).
\newblock Determining occurrence dynamics when false positives occur:
  {E}stimating the range dynamics of wolves from public survey data.
\newblock {\em PLoS One}, 8(6):e65808.

\bibitem[Miller et~al., 2011]{miller2011improving}
Miller, D.~A., Nichols, J.~D., McClintock, B.~T., Grant, E. H.~C., Bailey,
  L.~L., and Weir, L.~A. (2011).
\newblock Improving occupancy estimation when two types of observational error
  occur: {N}on-detection and species misidentification.
\newblock {\em Ecology}, 92(7):1422--1428.

\bibitem[Molinari, 2008]{molinari2008partial}
Molinari, F. (2008).
\newblock Partial identification of probability distributions with
  misclassified data.
\newblock {\em Journal of Econometrics}, 144(1):81--117.

\bibitem[Mulick et~al., 2022]{mulick2022bayesian}
Mulick, A.~R., Oza, S., Prieto-Merino, D., Villavicencio, F., Cousens, S., and
  Perin, J. (2022).
\newblock A {B}ayesian hierarchical model with integrated covariate selection
  and misclassification matrices to estimate neonatal and child causes of
  death.
\newblock {\em Journal of the Royal Statistical Society Series A: Statistics in
  Society}, 185(4):2097--2120.

\bibitem[Neelon, 2019]{neelon2019bayesian}
Neelon, B. (2019).
\newblock Bayesian zero-inflated negative binomial regression based on
  {P}{\'o}lya-{G}amma mixtures.
\newblock {\em Bayesian Analysis}, 14(3):829.

\bibitem[Nolan et~al., 2011]{nolan2011estimating}
Nolan, J.~J., Haas, S.~M., and Napier, J.~S. (2011).
\newblock Estimating the impact of classification error on the ``statistical
  accuracy''' of uniform crime reports.
\newblock {\em Journal of Quantitative Criminology}, 27:497--519.

\bibitem[Osborne et~al., 2019]{osborne2019utilizing}
Osborne, D.~L., Swartz, K., and Stover, A. (2019).
\newblock Utilizing the national incident-based reporting system to further our
  understanding of agricultural theft.
\newblock {\em International Journal of Rural Criminology}, 4(2):240--257.

\bibitem[Parker~Jr et~al., 2020]{parker2020species}
Parker~Jr, K.~A., Li, H., and Kalcounis-Rueppell, M.~C. (2020).
\newblock Species-specific environmental conditions for winter bat acoustic
  activity in {N}orth {C}arolina, {U}nited {S}tates.
\newblock {\em Journal of Mammalogy}, 101(6):1502--1512.

\bibitem[P{\'e}rez et~al., 2007]{perez2007misclassified}
P{\'e}rez, C.~J., Gir{\'o}n, F.~J., Mart{\'\i}n, J., Ruiz, M., and Rojano, C.
  (2007).
\newblock Misclassified multinomial data: {A} {B}ayesian approach.
\newblock {\em RACSAM}, 101(1):71--80.

\bibitem[Pina-S{\'a}nchez et~al., 2023]{pina2023impact}
Pina-S{\'a}nchez, J., Buil-Gil, D., Brunton-Smith, I., and Cernat, A. (2023).
\newblock The impact of measurement error in regression models using police
  recorded crime rates.
\newblock {\em Journal of Quantitative Criminology}, 39(4):975--1002.

\bibitem[Pocock et~al., 2014]{pocock2014choosing}
Pocock, M.~J., Chapman, D.~S., Sheppard, L.~J., and Roy, H.~E. (2014).
\newblock {\em Choosing and Using Citizen Science: a guide to when and how to
  use citizen science to monitor biodiversity and the environment}.
\newblock NERC/Centre for Ecology \& Hydrology.

\bibitem[Pollock et~al., 2018]{pollock2018madness}
Pollock, J., Glendinning, L., Wisedchanwet, T., and Watson, M. (2018).
\newblock The madness of microbiome: {A}ttempting to find consensus “best
  practice” for 16s microbiome studies.
\newblock {\em Applied and Environmental Microbiology}, 84(7):e02627--17.

\bibitem[Polson et~al., 2013]{polson2013bayesian}
Polson, N.~G., Scott, J.~G., and Windle, J. (2013).
\newblock Bayesian inference for logistic models using {P}{\'o}lya--{G}amma
  latent variables.
\newblock {\em Journal of the American Statistical Association},
  108(504):1339--1349.

\bibitem[Rodr{\'\i}guez-San~Pedro et~al., 2024]{rodriguez2024species}
Rodr{\'\i}guez-San~Pedro, A., Allendes, J.~L., Bruna, T., and Grez, A.~A.
  (2024).
\newblock Species-specific responses of insectivorous bats to weather
  conditions in central {C}hile.
\newblock {\em Animals}, 14(6):860.

\bibitem[Royle and Link, 2006]{royle2006generalized}
Royle, J.~A. and Link, W.~A. (2006).
\newblock Generalized site occupancy models allowing for false positive and
  false negative errors.
\newblock {\em Ecology}, 87(4):835--841.

\bibitem[Royle and Nichols, 2003]{royle2003estimating}
Royle, J.~A. and Nichols, J.~D. (2003).
\newblock Estimating abundance from repeated presence--absence data or point
  counts.
\newblock {\em Ecology}, 84(3):777--790.

\bibitem[Ruiz-Gutierrez et~al., 2016]{ruiz2016uncertainty}
Ruiz-Gutierrez, V., Hooten, M.~B., and Campbell~Grant, E.~H. (2016).
\newblock Uncertainty in biological monitoring: {A} framework for data
  collection and analysis to account for multiple sources of sampling bias.
\newblock {\em Methods in Ecology and Evolution}, 7(8):900--909.

\bibitem[Rydberg and Carkin, 2017]{rydberg2017utilizing}
Rydberg, J. and Carkin, D.~M. (2017).
\newblock Utilizing alternate models for analyzing count outcomes.
\newblock {\em Crime \& Delinquency}, 63(1):61--76.

\bibitem[Salda{\~n}a-V{\'a}zquez and Mungu{\'\i}a-Rosas,
  2013]{saldana2013lunar}
Salda{\~n}a-V{\'a}zquez, R.~A. and Mungu{\'\i}a-Rosas, M.~A. (2013).
\newblock Lunar phobia in bats and its ecological correlates: a meta-analysis.
\newblock {\em Mammalian Biology}, 78(3):216--219.

\bibitem[Scharf et~al., 2022]{scharf2022constructing}
Scharf, H.~R., Lu, X., Williams, P.~J., and Hooten, M.~B. (2022).
\newblock Constructing flexible, identifiable and interpretable statistical
  models for binary data.
\newblock {\em International Statistical Review}, 90(2):328--345.

\bibitem[Schaub and Abadi, 2011]{schaub2011integrated}
Schaub, M. and Abadi, F. (2011).
\newblock Integrated population models: a novel analysis framework for deeper
  insights into population dynamics.
\newblock {\em Journal of Ornithology}, 152:227--237.

\bibitem[Schmidt et~al., 2013]{schmidt2013site}
Schmidt, B.~R., K{\'e}ry, M., Ursenbacher, S., Hyman, O.~J., and Collins, J.~P.
  (2013).
\newblock Site occupancy models in the analysis of environmental {DNA}
  presence/absence surveys: {A} case study of an emerging amphibian pathogen.
\newblock {\em Methods in Ecology and Evolution}, 4(7):646--653.

\bibitem[Shuler et~al., 2021]{shuler2021bayesian}
Shuler, K., Verbanic, S., Chen, I.~A., and Lee, J. (2021).
\newblock A {B}ayesian nonparametric analysis for zero-inflated multivariate
  count data with application to microbiome study.
\newblock {\em Journal of the Royal Statistical Society: Series C (Applied
  Statistics)}, 70(4):961--979.

\bibitem[Skogan, 1974]{skogan1974validity}
Skogan, W.~G. (1974).
\newblock The validity of official crime statistics: {A}n empirical
  investigation.
\newblock {\em Social Science Quarterly}, pages 25--38.

\bibitem[Spiers et~al., 2022]{spiers2022estimating}
Spiers, A.~I., Royle, J.~A., Torrens, C.~L., and Joseph, M.~B. (2022).
\newblock Estimating species misclassification with occupancy dynamics and
  encounter rates: {A} semi-supervised, individual-level approach.
\newblock {\em Methods in Ecology and Evolution}, 13(7):1528--1539.

\bibitem[Steorts et~al., 2016]{steorts2016bayesian}
Steorts, R.~C., Hall, R., and Fienberg, S.~E. (2016).
\newblock A {B}ayesian approach to graphical record linkage and deduplication.
\newblock {\em Journal of the American Statistical Association},
  111(516):1660--1672.

\bibitem[Stratton, 2022]{Stratton2022}
Stratton, C. (2022).
\newblock Strattonch/{C}oupled{U}ncoupled: {C}oupling validation effort
  manuscript release (v1.0.0).
\newblock Zenodo. https://doi.org/10.5281/zenodo.6040068.

\bibitem[Stratton et~al., 2022]{stratton2022coupling}
Stratton, C., Irvine, K.~M., Banner, K.~M., Wright, W.~J., Lausen, C., and Rae,
  J. (2022).
\newblock Coupling validation effort with in situ bioacoustic data improves
  estimating relative activity and occupancy for multiple species with
  cross-species misclassifications.
\newblock {\em Methods in Ecology and Evolution}, 13(6):1288--1303.

\bibitem[Swartz et~al., 2004]{swartz2004bayesian}
Swartz, T.~B., Haitovsky, Y., Vexler, A., and Yang, T.~Y. (2004).
\newblock Bayesian identifiability and misclassification in multinomial data.
\newblock {\em Canadian Journal of Statistics}, 32(3):285--302.

\bibitem[Tancredi and Liseo, 2011]{tancredi2011hierarchical}
Tancredi, A. and Liseo, B. (2011).
\newblock A hierarchical {B}ayesian approach to record linkage and population
  size problems.
\newblock {\em The Annals of Applied Statistics}, 5(2B):1553--1585.

\bibitem[Thies et~al., 2006]{thies2006influence}
Thies, W., Kalko, E.~K., and Schnitzler, H.-U. (2006).
\newblock Influence of environment and resource availability on activity
  patterns of \textit{{C}arollia castanea} ({P}hyllostomidae) in {P}anama.
\newblock {\em Journal of Mammalogy}, 87(2):331--338.

\bibitem[Tyre et~al., 2003]{tyre2003improving}
Tyre, A.~J., Tenhumberg, B., Field, S.~A., Niejalke, D., Parris, K., and
  Possingham, H.~P. (2003).
\newblock Improving precision and reducing bias in biological surveys:
  {E}stimating false-negative error rates.
\newblock {\em Ecological Applications}, 13(6):1790--1801.

\bibitem[V{\'a}squez et~al., 2020]{vasquez2020species}
V{\'a}squez, D.~A., Grez, A.~A., and Rodr{\'\i}guez-San~Pedro, A. (2020).
\newblock Species-specific effects of moonlight on insectivorous bat activity
  in central {C}hile.
\newblock {\em Journal of Mammalogy}, 101(5):1356--1363.

\bibitem[Voigt et~al., 2011]{voigt2011rain}
Voigt, C.~C., Schneeberger, K., Voigt-Heucke, S.~L., and Lewanzik, D. (2011).
\newblock Rain increases the energy cost of bat flight.
\newblock {\em Biology Letters}, 7(5):793--795.

\bibitem[Wadsworth et~al., 2017]{wadsworth2017integrative}
Wadsworth, W.~D., Argiento, R., Guindani, M., Galloway-Pena, J., Shelburne,
  S.~A., and Vannucci, M. (2017).
\newblock An integrative {B}ayesian {D}irichlet-multinomial regression model
  for the analysis of taxonomic abundances in microbiome data.
\newblock {\em BMC Bioinformatics}, 18(1):94.

\bibitem[Wang et~al., 2020]{wang2020inference}
Wang, S., Wang, L., and Swartz, T.~B. (2020).
\newblock Inference for misclassified multinomial data with covariates.
\newblock {\em Canadian Journal of Statistics}, 48(4):655--669.

\bibitem[Wheeler and Kovandzic, 2018]{wheeler2018monitoring}
Wheeler, A.~P. and Kovandzic, T.~V. (2018).
\newblock Monitoring volatile homicide trends across {US} cities.
\newblock {\em Homicide Studies}, 22(2):119--144.

\bibitem[Willoughby et~al., 2016]{willoughby2016importance}
Willoughby, J.~R., Wijayawardena, B.~K., Sundaram, M., Swihart, R.~K., and
  DeWoody, J.~A. (2016).
\newblock The importance of including imperfect detection models in {eDNA}
  experimental design.
\newblock {\em Molecular Ecology Resources}, 4(16):837--844.

\bibitem[Wormeli, 2018]{wormeli2018criminal}
Wormeli, P. (2018).
\newblock Criminal justice statistics — {A}n evolution.
\newblock {\em Criminology \& Public Policy}, 17(2):483--496.

\bibitem[Wright et~al., 2020]{wright2020modelling}
Wright, W.~J., Irvine, K.~M., Almberg, E.~S., and Litt, A.~R. (2020).
\newblock Modelling misclassification in multi-species acoustic data when
  estimating occupancy and relative activity.
\newblock {\em Methods in Ecology and Evolution}, 11(1):71--81.

\bibitem[Xu et~al., 2015]{xu2015assessment}
Xu, L., Paterson, A.~D., Turpin, W., and Xu, W. (2015).
\newblock Assessment and selection of competing models for zero-inflated
  microbiome data.
\newblock {\em PloS One}, 10(7):e0129606.

\bibitem[Zhang and Yi, 2020]{zhang2020nbzimm}
Zhang, X. and Yi, N. (2020).
\newblock {NBZIMM:} {N}egative binomial and zero-inflated mixed models, with
  application to microbiome/metagenomics data analysis.
\newblock {\em BMC Bioinformatics}, 21(1):1--19.

\end{thebibliography}

\end{document}